\documentclass[a4paper,11pt]{article}
\pdfoutput=1 

\usepackage{jcappub} 

\usepackage[T1]{fontenc} 

\usepackage[sc]{mathpazo}

\newcommand{\mWIMP}{m_{\chi}}
\newcommand{\mNucleus}{m_{N}}
\newcommand{\rNucleus}{r_{N}}
\newcommand{\sigmaWIMPNucleus}{\sigma_{\chi-N}}
\newcommand{\sigmaWIMPNucleon}{\sigma_{n}}

\begin{document}
\title{Directional dark matter detection sensitivity of a two-phase liquid argon detector}

\newcommand{\Cagliari}{Physics Department, Universit\`a degli Studi, Cagliari 09042, Italy}
\newcommand{\CagliariINFN}{Istituto Nazionale di Fisica Nucleare, Sezione di Cagliari, Cagliari 09042, Italy}
\newcommand{\Pisa}{Physics Department, Universit\`a degli Studi di Pisa, Pisa 56127, Italy}
\newcommand{\PisaINFN}{Istituto Nazionale Fisica Nucleare, Sezione di Pisa, Pisa 56127, Italy}
\newcommand{\Genova}{Physics Department, Universit\`a degli Studi di Genova, Genova 16146, Italy}
\newcommand{\GenovaINFN}{Istituto Nazionale di Fisica Nucleare, Sezione di Genova, Genova 16146, Italy}
\newcommand{\LNGS}{INFN Laboratori Nazionali del Gran Sasso, Assergi (AQ) 67010, Italy}
\newcommand{\Napoli}{Physics Department, Universit\`a degli Studi Federico II and INFN, Napoli 80126, Italy}
\newcommand{\NapoliINFN}{Istituto Nazionale di Fisica Nucleare, Sezione di Napoli, Napoli 80126, Italy}
\newcommand{\GSSI}{Gran Sasso Science Institute, L'Aquila AQ 67100, Italy}
\newcommand{\Roma}{Physics Department, Sapienza Universit\`a di Roma, Roma 00185, Italy}
\newcommand{\RomaINFN}{Istituto Nazionale di Fisica Nucleare, Sezione di Roma 1, Roma 00185, Italy}
\newcommand{\Princeton}{Physics Department, Princeton University, Princeton, NJ 08544, USA}
\newcommand{\Enna}{Universit\`a di Enna KORE, Enna 94100, Italy}
\newcommand{\CataniaINFN}{Istituto Nazionale Fisica Nucleare, Laboratori Nazionali del Sud, 95123  Catania, Italy}
\newcommand{\MilanoINFN}{Istituto Nazionale di Fisica Nucleare, Sezione di Milano, Milano 20133, Italy}
\newcommand{\Paris}{APC, Universit\'e Paris Diderot, CNRS/IN2P3, CEA/Irfu, Obs. de Paris, Sorbonne Paris Cit\'e, Paris 75205, France}
\newcommand{\LPNHE}{LPNHE Paris, Universit\'e Pierre et Marie Curie, Universit\'e Paris Diderot, CNRS/IN2P3, Paris 75252, France}
\newcommand{\Houston}{Department of Physics, University of Houston, Houston, TX 7704, USA}

\author[a,b,1]{M. Cadeddu,\note{Corresponding author.}}
\author[b]{M. Lissia,}
\author[c,d]{P. Agnes,}
\author[e,f]{G. Batignani,}
\author[b]{W. M. Bonivento,}
\author[g,h]{B. Bottino,}
\author[a,b]{M. Caravati,}
\author[i,l]{S. Catalanotti,}
\author[i,l]{V. Cataudella,}
\author[b]{C.~Cical\`{o},}
\author[l]{A. Cocco,}
\author[i,l]{G. Covone,}
\author[i,l]{A. de Candia,}
\author[i,l]{G. De Filippis,}
\author[i,l]{G. De Rosa,}
\author[h,m,n]{S. Davini,}
\author[a,b]{A. Devoto,}
\author[o,p]{C. Dionisi,}
\author[c]{D. Franco,}
\author[q]{C. Giganti,}
\author[r,s]{C. Galbiati,}
\author[o,p]{S. Giagu,}
\author[t]{M. Gulino,}
\author[f]{M. Kuss,}
\author[l]{L. Lista,}
\author[i,l]{G. Longo,}
\author[q]{A. Navrer-Agasson,}
\author[g,h]{M. Pallavicini,}
\author[u]{L. Pandola,}
\author[e,f]{E. Paoloni,}
\author[a]{E. Picciau,}
\author[b]{M. Razeti,}
\author[p]{M. Rescigno,}
\author[c]{Q. Riffard,}
\author[l,r]{B. Rossi,}
\author[n]{N. Rossi,}
\author[h]{G. Testera,}
\author[i,l]{P. Trinchese,}
\author[c]{A. Tonazzo,}
\author[i,l]{S. Walker,}

\author[i,l]{and G. Fiorillo}


\affiliation[a]{\Cagliari}
\affiliation[b]{\CagliariINFN}
\affiliation[c]{\Paris}
\affiliation[d]{\Houston}
\affiliation[e]{\Pisa}
\affiliation[f]{\PisaINFN}
\affiliation[g]{\Genova}
\affiliation[h]{\GenovaINFN}
\affiliation[i]{\Napoli}
\affiliation[l]{\NapoliINFN}
\affiliation[m]{\GSSI}
\affiliation[n]{\LNGS}
\affiliation[o]{\Roma}
\affiliation[p]{\RomaINFN}
\affiliation[q]{\LPNHE}
\affiliation[r]{\Princeton}
\affiliation[s]{\MilanoINFN}
\affiliation[t]{\Enna}
\affiliation[u]{\CataniaINFN}

\emailAdd{matteo.cadeddu@ca.infn.it}

\abstract{
We examine the sensitivity of a large scale two-phase liquid argon detector to the directionality of the dark matter signal. 
This study was performed under the assumption that, above 50 keV of recoil energy, one can determine (with some resolution) the direction of the recoil nucleus without head-tail discrimination, as suggested by past studies that proposed to exploit the dependence of columnar recombination on the angle between the recoil nucleus direction and the electric field.
In this paper we study the differential interaction recoil rate as a function of the recoil direction angle with respect to the zenith for a detector located at the Laboratori Nazionali del Gran Sasso and we determine its diurnal and seasonal modulation. Using a likelihood-ratio based approach we show that, with the angular information alone, 100 (250) events are enough to reject the isotropic hypothesis at three standard deviation level, for a perfect (400~mrad) angular resolution. For an exposure of 100 tonne years this would correspond to a spin independent WIMP-nucleon cross section of  about $\textrm{10}^{-46}\textrm{\ cm}^{2}$
at 200 GeV WIMP mass. The results presented in this paper provide strong motivation for the experimental determination of directional recoil effects in two-phase liquid argon detectors.}
\date{29/06/2018}

\maketitle
\flushbottom

\section{Introduction}
\label{sec:intro}
Dark Matter (DM) is the most compelling indirect evidence for physics beyond the Standard Model.
It constitutes about 80\% of the mass of the Universe~\cite{Ade:2016bk} and has a fundamental role in
the comprehension of the evolution of the Universe since the Big Bang.
Understanding the nature of DM is therefore one of the most intriguing puzzles in physics.
In the last decades many measurements have been gathered that could be 
explained by assuming a large amount of  DM in cosmic structures at different mass scales.
Starting from rotation curves of spiral galaxies~\cite{Rubin:1999hb},
the most persuasive evidences come from the observations of
anisotropies of the Cosmic Microwave Background~\cite{Ade:2016bk}, 
gravitational lensing on galaxy clusters~\cite{Clowe:2006hr} 
and galaxy scales~\cite{Covone:2009ji}, and the Big-Bang nucleosynthesis~\cite{Malaney:1993ep}.
A plausible model for DM is that it consists of non-relativistic
Weakly-Interacting Massive Particles (WIMPs). In our Galaxy, the density
distribution of this DM halo, which extends far beyond
the visible disk, are inferred from the rotation curves of the visible
matter~\cite{Nesti:2013jc,Catena:2010fv}.  The velocity distribution is less understood but it can be inferred using different approaches~\cite{Sloane:2016kyi,Herzog-Arbeitman:2017fte,Necib:2018iwb}.

Direct detection DM experiments
look for possible WIMP interactions with target nuclei aiming to detect an 
excess of WIMP-induced nuclear recoils above the estimated 
background~\cite{Bertone:2005bi}. Establishing this excess is a serious experimental challenge,
given the expected electron and neutron backgrounds that mimic nuclear
recoils from WIMPs and the low expected rate. It is necessary to remove
radioactive backgrounds to a technologically-challenging low level
and to rely on the detector capability to discriminate the remaining
backgrounds. A non-directional detector could enhance the signal over
background ratio using the expected annual modulation of the DM signal
due to the Earth motion around the Sun~\cite{Drukier:1986kv,Freese:1987wu}.
For instance the DAMA collaboration~\cite{Bernabei:2008ik, Bernabei:2018yyw} reported an observation
of such a modulation.
However, this seasonal modulation is expected to be smaller than 10\% and 
background sources exist that have similar seasonal modulations~\cite{Tiwari:2017xnj}.

A large mass  detector with sensitivity to the direction of the recoiling nuclei
would constitute a considerable breakthrough in the search for DM, as we shall argue in this work. 
A directional detector would allow one to prove that the detected new particle is indeed a 
dark matter candidate.

For the sake of concreteness  a detector located at the latitude\footnote{The LNGS coordinates are 42\textdegree~28' N 13\textdegree~33' E.}
of the INFN Laboratori Nazionali del Gran Sasso (LNGS), Italy, 
where the  DarkSide-20k experiment~\cite{Aalseth:2017fik} will be located, is considered.

We show that the expected event rate varies by a large factor ($4-8$)
when considering nuclear recoil directions going from the zenith to the
horizon and, at fixed angular direction, it varies by about the same factor
with sidereal-day period.
The angular resolution of the detector will imply important consequences on
the experimental sensitivity to such a rate variation.
The event-rate variations as a function of the sidereal time
and as a function of the polar angle
are very robust and are largely independent on details of the WIMP interaction
and of the WIMP velocity distribution: they are direct consequences
of the solar system motion through the Galaxy, the Earth revolution around the Sun and its rotation. Isotropic backgrounds (\textit{e.g.:} Diffuse supernovae
and atmospheric neutrinos), backgrounds from sources within the solar system (\textit{e.g.:}
solar neutrinos), or backgrounds with the periodicity of solar day
(\textit{e.g.:} backgrounds that depend on the temperature or the atmospheric
density) can be considerably reduced using the angular and time information provided by a directional detector.

Several prototypes of directional detectors exist~\cite{Mayet:2016zxu,2016PhR...662....1B}, generally based on the attempt to perform an imaging of the nuclear recoil trajectory. These detectors aim at achieving high spatial resolutions and are usually limited in mass, thus being capable to collect limited exposures. 
On the other hand, non-directional DM detectors have already reached exposures greater than $\textrm{10}^{5}$~kg-day, \cite{Aprile:2018dbl} excluding spin independent WIMP-nucleon cross section lower than about $\textrm{1.7}\times\textrm{10}^{-46}\textrm{\ cm}^{2}$ for a 200 GeV WIMP mass.

As argued in~\cite{Nygren:2013fy}, a promising technique for a very large-mass detector with directional DM capability would be to exploit the phenomenon called \emph{Columnar Recombination} (CR) in a noble liquid Time Projection Chamber (TPC). An argon-based detector sensitive to the effect of CR, would combine directional sensitivity with the ability to collect exposures of several hundreds of tonnes year ~\cite{Aalseth:2017fik}.

In noble liquid TPCs the recoiling nucleus produces both scintillation and
ionization. CR models~\cite{Jaffe:1929gz, Cadeddu2017Re} predict that the amount of signal due to ionization that can be collected
in the presence of an electric field ${\vec{\cal E}}$ should depend on the angle
$\theta_r$ between $\vec{\cal E}$ and the track (the average direction of the straggling nucleus). The ionization signal is expected to be
maximal when $\theta_r=90^{\circ}$, since electrons drift
in a direction perpendicular to the region around the recoil track
where ions are present, minimizing recombination. On the contrary, it should be minimal when $\theta_r=0^{\circ}$,
since electrons drift along the region where ions are present, with high probability
of recombination. The ionization signal from the collected electrons
would be a function of the component of the electric field perpendicular
to the track, ${\cal E}_{\perp}={\cal E}\sin\theta_r$, and, therefore
would carry, together with the scintillation signal, information
on the average direction of the recoiling nucleus. 
CR in a Liquid Argon (LAr) TPC would thus provide signatures for the orientation of the ionizing tracks relative to the direction of electric field. Evidence for this effect has been collected for $\alpha$ particles and protons~\cite{Swan:1965ha, Acciarri:2010gm} with energies of 5.14 MeV and between 50-250 MeV, respectively.  The SCENE experiment~\cite{PhysRevD.88.092006, Cao:2015ks}, a small two-phase LAr TPC designed for calibration of nuclear-recoil responses, gave a hint for the same directional signature in the scintillation response of nuclear recoils of about 57 keV, 
approximately the energy at which, following the argument in~\cite{Nygren:2013fy}, one might expect the ion range to be sufficient  to form a track with a definite direction.

General aspects of DM directional detection  have
been discussed in a number of papers~\cite{Billard:2012fia,Bozorgnia:2012kq,Copi:2001kz,Mayet:2016zxu}.
In the following, an active mass of 100 tonne (which in terms of number of WIMP events is equivalent to a 20 tonne active mass detector running for 5 years)
is considered with a detector at LNGS
as in the DarkSide-20k experiment~\cite{Aalseth:2017fik}. Namely, DarkSide-20k will be able to collect such exposure keeping the number of instrumental background interactions to less than 0.1 events. At the location of the laboratory, the angle between the expected average WIMP direction 
with the vertical electric field in the LAr TPC spans the entire range 
between 0\textdegree~and 90\textdegree~during the day.
Preliminary results have been presented in conference 
proceedings~\cite{Cadeddu:2016mac, Cadeddu:2017zad,Fiorillo:2017kgd}, but
the statistical analysis, seasonal modulation and the application to larger detectors are shown
for the first time in this paper.

The paper is outlined as follows.
Section~\ref{sec:RecoilRates} reviews the theoretical
framework, introducing the formulae for the recoil cross section and rates as a function
of the relevant angular variable, velocity distribution and coordinate
systems.
Section~\ref{sec:DirectionalRecoilRateLNGS} presents
the  recoil angular distributions,
while Section~\ref{sec:SeasonalEffects} discusses the annual modulation of the signals.
Section~\ref{sec:statanal} presents a simplified statistical analysis
method to study the DM directionality.
We draw our conclusions in Section~\ref{sec:conclusions}.  

\section{Recoil rates}
\label{sec:RecoilRates} 

\subsection{Cross section and differential rates}
\label{subsec:Kinematic} 

\begin{figure}[htbp]
\centering{} \includegraphics[angle=0,width=4.8cm]{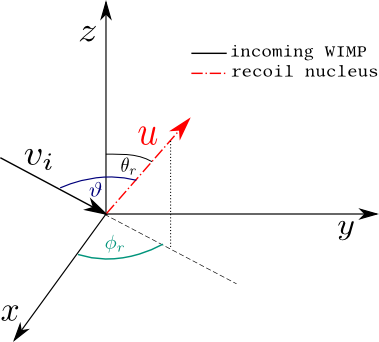}
\protect\protect\caption[aa]{\label{fig:RecoilAnglesSdr}
  Schematic view of a WIMP-nucleus scattering in a reference frame such that $x$, $y$ and $z$ are laboratory-fixed coordinates.
The incoming WIMP with velocity $\mathbf{v}_i$ hits the nucleus that recoils in the direction of the momentum $\mathbf{q}=m_N\mathbf{u}$ whose  azimuthal and polar angles are  $\phi_{r}$
and $\theta_{r}$. The angle between $u$ and $\mathbf{v}_i$ is $\vartheta$.}
 \end{figure}

In a given reference frame, let's assume $\mathbf{v}_i$ is the velocity of the incoming
WIMP of mass $\mWIMP$,
$\mathbf{u}$ is the velocity of the recoiling nucleus of mass
$\mNucleus$, $\mathbf{q}=\mNucleus\mathbf{u}$
is the nucleus momentum and $E_{r}=q^{2}/(2\mNucleus)$ is the corresponding energy.
The azimuthal and polar angles of the recoiling nucleus are $\phi_{r}$
and $\theta_{r}$, while $\vartheta$ is
the angle between the incoming WIMP direction and
the recoiling nucleus, as shown in Fig.~\ref{fig:RecoilAnglesSdr}.
In general, recoil rates are convolutions of the scattering cross section
and the incoming velocity distribution. A point-like cross section in the center of mass is constant and can
be pa\-ra\-me\-tri\-zed by the total WIMP-nucleus cross section
$\sigmaWIMPNucleus$. For a spin-independent interaction with equal couplings
for neutrons and protons, $\sigmaWIMPNucleus$ can be expressed in terms of WIMP-nucleon cross section $\sigmaWIMPNucleon$ as
$\sigmaWIMPNucleus /\mu_{N}^{2} = A^{2}\sigmaWIMPNucleon /\mu_{p}^{2}$, where $A$
is the atomic mass and $\mu_{N}$ and $\mu_{p}$ are the WIMP-nucleus
and the WIMP-nucleon reduced masses, respectively.
The finite size of the nucleus is taken into account by introducing the Helm
nuclear form factor~\cite{Helm:1956dv}:
\begin{equation}
F\left(q\right)=
\frac{3\left[\sin\left(q\rNucleus\right)-q\rNucleus\cos\left(q\rNucleus\right)\right]}{\left(q\rNucleus\right)^{3}}
e^{-\left(qs\right)^{2}/2}\,,\label{eq:HelmFormFactor}
\end{equation}
where $s=0.9$ fm is the surface thickness and $r_{N} = 3.9\,\,\mathrm{fm}$ is the argon nucleus effective radius.

In the laboratory frame, where the target nucleus is at rest, the double-differential
cross section depends on the cosine of the angle between
the incoming WIMP and the recoiling nucleus $\hat{\mathbf{v}}_{i}\cdot\hat{\mathbf{q}}=\cos\vartheta$ as
\begin{eqnarray}
\frac{\mathrm{d}^{2}\sigma(q,\hat{\mathbf{v}}_{i}\cdot\hat{\mathbf{q}})}{\mathrm{d}q^{2}\,d\varOmega}
& = &\frac{\mathrm{d}^{2}\sigma(q,\cos\vartheta)}{2\mNucleus\,\mathrm{d}E_{r}2\pi\, \mathrm{d}\cos\vartheta}
\\
 & = &\frac{\sigmaWIMPNucleus}{8\pi\mu_{N}^{2}v_{i}}\:F^{2}\left(q\right)
\delta\left(\mathbf{v}_{i}\cdot\hat{\mathbf{q}}-\frac{q}{2\mu_{N}}\right)\,.\nonumber
\label{eq:DoubleDiffCrossSection}
\end{eqnarray}

Given a velocity distribution for the incoming WIMP $f(\mathbf{v}_{i})$,
normalized so that $\int f(\mathbf{v})\mathrm{d}\mathbf{v}=1$, and a WIMP
mass density $\rho$, the double-differential recoil rate per unit
mass, \textit{i.e.} the rate per target nucleus divided by the nucleus mass $\mNucleus$,
as a function of the nuclear recoil energy, $E_{r}$, and of the recoil
direction $\hat{\mathbf{q}}$ is 
\begin{eqnarray}
  \frac{\mathrm{d}^{2}R(E_{r},\hat{\mathbf{q}})}{\mathrm{d}E_{r}\,\mathrm{d}\Omega_{r}} 
& = & \frac{2\rho}{\mWIMP}\int v\frac{\mathrm{d}^{2}\sigma(q,\hat{\mathbf{v}}\cdot\hat{\mathbf{q}})}{\mathrm{d}q^{2}\,\mathrm{d}\varOmega}
       \,f\left(\mathbf{v}\right)\,\mathrm{d}\mathbf{v}\\
 & = & \frac{\rho \,\sigmaWIMPNucleus F^{2}(q)}{\mWIMP 4\pi\,\mu_{N}^{2}}\int
       \delta\left(\mathbf{v}\cdot\hat{\mathbf{q}}-\frac{q}{2\mu_{N}}\right)f\left(\mathbf{v}\right)\mathrm{d}\mathbf{v}\nonumber\\
 & = & \frac{\rho}{\mWIMP}\,\frac{\sigmaWIMPNucleus F^{2}(q)}{4\pi\mu_{N}^{2}}
 \hat{f}\left(v_{min},\hat{\mathbf{q}}\right)\,,\nonumber\label{eq:DoubleDiffRate}
\end{eqnarray}
where $v_{min}=q/(2\mu_{N})=\sqrt{2\mNucleus E_{r}}/(2\mu_{N})$ is the minimal
WIMP velocity that can give momentum $q$ or energy $E_{r}$ to the
recoiling nucleus and $\hat{f}\left(v_{min},\hat{\mathbf{q}}\right)$
is the 3-dimensional Radon transform \cite{Gondolo:2002fg} of the velocity distribution
$f(\mathbf{v})$.

In this paper we assume the Standard Halo Model (SHM), \textit{i.e.}
an isotro\-pic Maxwell-Boltzmann  WIMP velocity distribution of
width $\sigma_{v}$ in a reference frame at rest with respect
to the Galactic center\footnote{The usage of a truncated velocity distribution at the escape velocity of 544 km/s (see Ref.~\cite{Lisanti:2010qx} for details) has a negligible impact on the event rate and on the rate distributions.}. In a reference frame with velocity $\mathbf{V}$
relative to the Galactic center, the velocity distribution is 
\begin{equation}
f(\mathbf{v})=\frac{1}{\sqrt{\left(2\pi\sigma_{v}^{2}\right)^{3}}}\exp\left[-\frac{1}{2}\left(\frac{\mathbf{v}+\mathbf{V}}{\sigma_{v}}\right)^{2}\right]\label{eq:SHMvelocityDistribution}
\end{equation}
and the corresponding Radon transform is 
\begin{equation}
\hat{f}\left(v_{min},\hat{\mathbf{q}}\right)=\frac{1}{\sqrt{2\pi\sigma_{v}^{2}}}
\exp\left[-\frac{1}{2}\left(\frac{v_{min}+\hat{\mathbf{q}}\cdot\mathbf{V}}{\sigma_{v}}\right)^{2}\right]
\,.\label{eq:SHMvelocityRadonTransform}
\end{equation}
Therefore, if recoils are measured in a frame at rest with respect
to the center of the Galaxy, $\mathbf{V}=0$ and the rate is isotropic. Similarly,
when measured in a frame at rest with respect to the Sun, $\mathbf{V}$
is the Sun velocity relative to the galactic center $\mathbf{V}_{SG}$,
which points towards the galactic coordinates~\cite{Blaauw:1960km}
($\ell_{c}$=90\textdegree, $\mathit{b_{c}}$=0\textdegree), roughly the direction of the Cygnus
constellation, and has magnitude $V_{SG}\approx v_{0}=220$ km/s,
where $v_{0}$ is the Galactic orbital speed at the Sun position.
For an Earthbound laboratory the velocity $\mathbf{V}$ can be decomposed
as $\mathbf{V}=\mathbf{V}_{SG}+\mathbf{V}_{ES}$, where $\mathbf{V}_{ES}$
is the Earth velocity relative to the Sun, which has magnitude $V_{ES}\approx30$
km/s, about ten times smaller than $v_{0}$. The laboratory speed
relative to the Earth center has been neglected, since it is almost
two orders of magnitude smaller than $\mathbf{V}_{ES}$ and the contribution to the rate is negligible. Clearly, its direction is accounted for being the crucial ingredient for the directional detection.

If a detector collects events of energy $E_{th}<E_{r}<E_{max}$, the
direction-dependent recoil rate per unit mass, obtained by substituting
the Radon transform from Eq.~(\ref{eq:SHMvelocityRadonTransform}) in
Eq.~(\ref{eq:DoubleDiffRate}) and integrating over the energy range, becomes 
\begin{equation}
\frac{\mathrm{d}R(E_{th},E_{max},\hat{\mathbf{q}})}{\mathrm{d}\Omega_{r}}
=\intop_{E_{th}}^{E_{max}}\mathrm{d}E_{r}\frac{\mathrm{d}^{2}R(E_{r},\hat{\mathbf{q}})}{\mathrm{d}E_{r}\,\mathrm{d}\Omega_{r}}
\, .\label{eq:RecoilEnergyIntegrated}
\end{equation}

The study presented in this paper focuses on the use of polar detectors \textit{i.e.} detectors
that give only information on the angle $\theta_{r}$ between the
recoil track and a fixed axis. If the fixed axis is the vertical direction, which corresponds to the z-axis as in Fig.~\ref{fig:RecoilAnglesSdr}, the relevant recoil rate is 
\begin{equation}
\frac{\mathrm{d}R(E_{th},E_{max},\cos\theta_{r})}{\mathrm{d}\cos\theta_{r}}
=\intop_{0}^{2\pi}\mathrm{d}\phi_{r}\intop_{E_{th}}^{E_{max}}\,\mathrm{d}E_{r}\:\frac{\mathrm{d}^{2}R(E_{r},\hat{\mathbf{q}})}{\mathrm{d}E_{r}\,\mathrm{d}\Omega_{r}}
\,,\label{eq:polarRate}
\end{equation}
which, after integrating out $\phi_{r}$, depends on $\cos\theta_{r}$.
In addition, if a detector cannot distinguish signals from recoil
tracks differing by 180$^{\circ}$, events that differ
by 180$^{\circ}$ are summed together. The relevant rate is the so-called
``folded'' angular recoil rate \cite{Gondolo}:
\begin{equation}
\frac{\mathrm{d}R_{F}(|\cos\theta_{r}|)}{\mathrm{d}|\cos\theta|}\equiv\frac{\mathrm{d}R}{\mathrm{d}\cos\theta_{r}}(\cos\theta_{r})
+\frac{\mathrm{d}R}{\mathrm{d}\cos\theta_{r}}(-\cos\theta_{r})\,,\label{eq:FoldedRate}
\end{equation}
which depends only on $|\cos\theta_r|$. Dependences of the recoil
rates on other variables are not shown.

Unless explicitly stated, this work shows results for a LAr detector
using the reference
values $\mWIMP=200$ GeV, $\mNucleus=0.923\,A$, where A is the argon atomic
mass, $\rho=0.3$ GeV cm$^{-3}$, and $\sigma_{v}=v_{0}/\sqrt{2}$.
Rates are given for  a reference cross section $\sigmaWIMPNucleon =\textrm{10}^{-46}\textrm{\ cm}^{2}$, which
is of the order of the last limits set by the LUX, Xenon1T and PANDAX-II collaborations \cite{Akerib:2017kg, Aprile:2017iyp, Cui:2017nnn}, for recoil energies
from $E_{th}=50\,\textrm{\ keV}$ to $E_{max}=200\textrm{\ keV}$, and for
an active mass of 100 tonne\footnote{This choice for the threshold energy is motivated by hints 
from the SCENE experiment~\cite{Cao:2015ks} for directional dependence in the scintillation signal 
at energy of 57.3 keV.}. 
Note that the anisotropy of all rates in Eqs.~(\ref{eq:RecoilEnergyIntegrated}),
(\ref{eq:polarRate}), and~(\ref{eq:FoldedRate}) depends only on
the velocity $\mathbf{V}$. In a given frame, which fixes $\mathbf{V}$,
one can choose different angular coordinate systems. If the angular coordinate system
is time dependent,\textit{ e.g.} a coordinate system fixed to the rotating Earth,
the direction of   $\mathbf{V}$ in that system becomes time dependent. In
a frame at rest with respect to the Earth and using Galactic coordinates,
$\mathbf{V}_{SG}$ is constant and only $\mathbf{V}_{ES}$ rotates
with the annual periodicity of the Earth revolution. Since $\mathbf{V}_{ES}$
is an order of magnitude smaller than $\mathbf{V}_{SG}$, 
the WIMP apparent direction  $-\mathbf{V}= -(\mathbf{V}_{SG} +\mathbf{V}_{ES} )$
rotates with annual periodicity around the fixed $\mathbf{V}_{SG}$
direction  with an opening angle of
about one tenth of radian. In this frame the peaked angular distribution is the main
signature of the signal and allows for background reduction. In the laboratory coordinate
system, the coordinates and, therefore, the apparent direction of $\mathbf{V}$ makes an additional rotation with
the periodicity of a sidereal day and an amplitude that depends on
the latitude. This specific periodicity is also a characteristic signature
and provides more background suppression.

\subsection{Recoil rate in Galactic angular coordinates}

\begin{figure}[htbp]
\centering{}\includegraphics[angle=0,width=7.5cm]{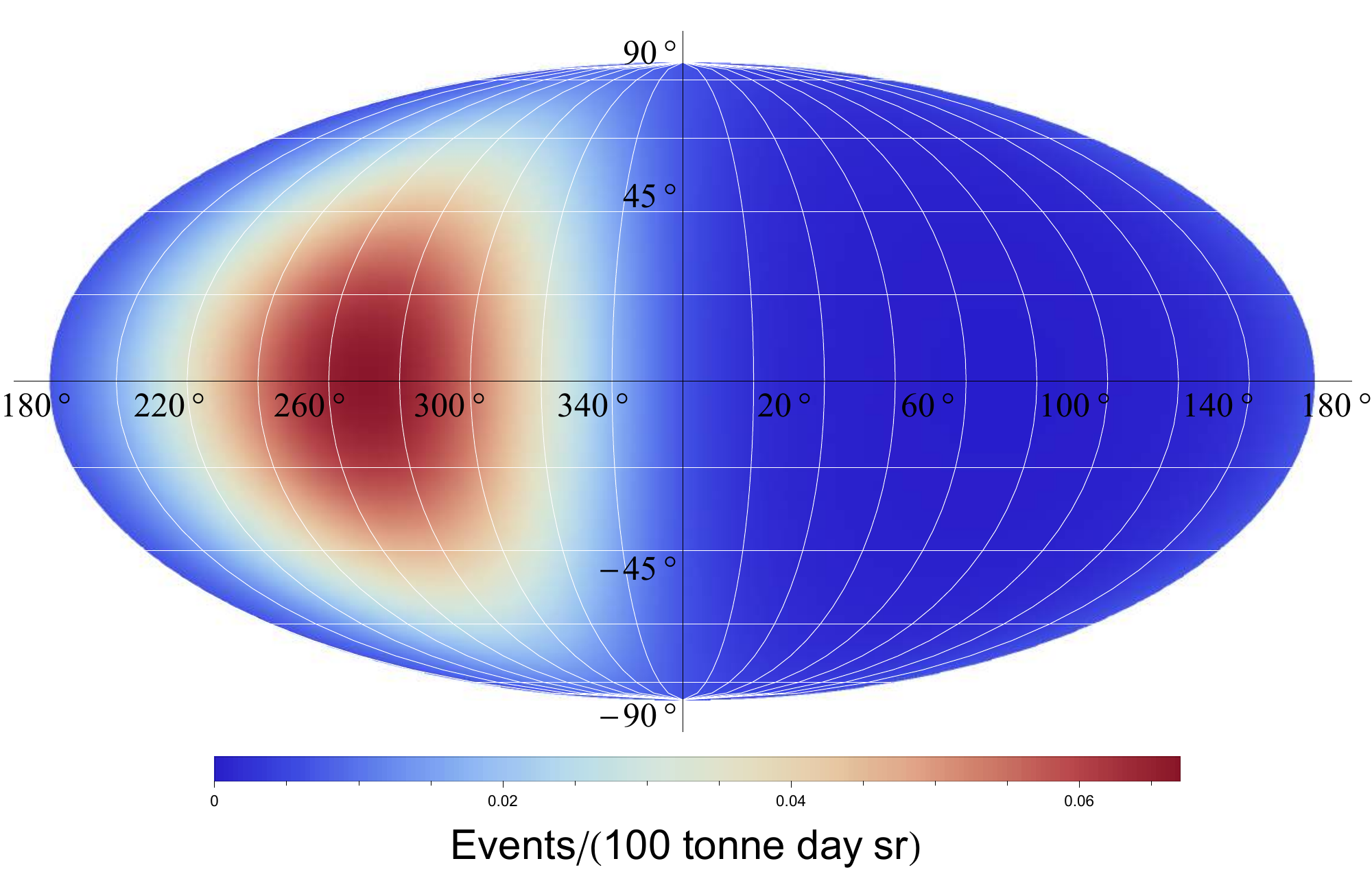} 
\protect\caption[bb]{\label{fig:mollweideRecoil} Recoil rate in argon, Eq~(\ref{eq:RecoilEnergyIntegrated}),
on a Mollweide equal area projection map of the celestial sphere in
galactic coordinates. The horizontal axis is the galactic longitude $0^{\circ}<\ell<360^{\circ}$
and the vertical axis is galactic latitude $-90^{\circ}<b<90^{\circ}$. The WIMP mass is 200 GeV, the WIMP-nucleon cross section
$\textrm{10}^{-46}\textrm{cm}^{2}$ and the energy interval $(50\mathrm{\ keV}\le E_r \le 200\textrm{\ keV})$.
The color scale is units of events/(100 tonne $\cdot$ day $\cdot$ sr).}
\end{figure}
\begin{figure}[htbp]
\centering{}\includegraphics[angle=0,width=8cm]{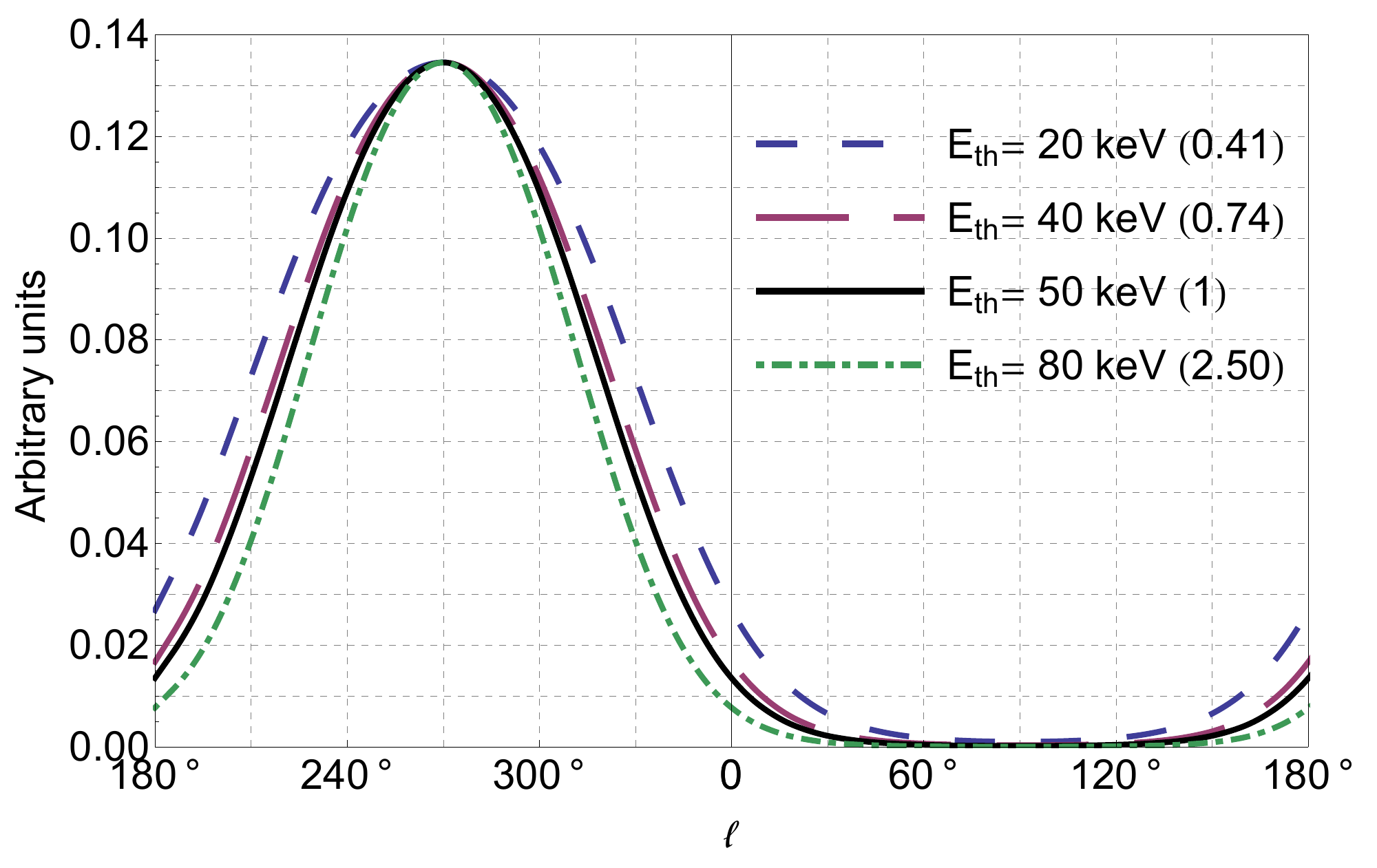}
\protect\caption[cc]{\label{fig:resolutionEth} 
Recoil rate in arbitrary units for argon 
as a function of the galactic longitude $\ell$ in degrees. The WIMP
mass is assumed to be 200 GeV and the recoil energy is integrated between $E_{th}$ and 200~keV, where
$E_{th}=$ 20, 40, 50 and 80 keV. With respect to the curve for $E_{th}=50$ keV the other curves have been rescaled
to have the same maximum. As shown in the legend the rescaling factors
are: 0.41, 0.74, 2.50, respectively. 
}
\end{figure}

 Figures
 \ref{fig:mollweideRecoil} and \ref{fig:resolutionEth}
show results in a reference frame at rest
with respect to the Sun in Galactic coordinates to demonstrate
the  potentialities of a directional detector independently of the location, as it has been extensively
done in the past, 
and to discuss the influence of the threshold energy $E_{\mathrm{th}}$.
All other results will be given for a detector located at the latitude
of LNGS in the local coordinate system with the polar axis pointing
in the vertical direction. Indeed, the potentialities of a directional detector and, more specifically,
the signature in the angular recoil rate of the detector motion through
the WIMP halo are best illustrated in Galactic coordinates in
a frame at rest with the Sun.
In this coordinate system, $\hat{x}$ points from the Sun towards
the Galactic center, $\hat{y}$ in the direction of the Solar motion
and $\hat{z}$ towards the Galactic north pole; therefore, $\mathbf{V}=v_{0}\hat{y}$.
In Figure \ref{fig:mollweideRecoil} we show the angular recoil rate
of Eq.~(\ref{eq:RecoilEnergyIntegrated}) for argon on a Mollweide equal area
projection map. The horizontal  axis is the galactic longitude $0^{\circ}<l<360^{\circ}$
(the counterclockwise angle from the $\hat{x}$ axis) and the vertical
axis is the galactic latitude $-90^{\circ}<b<90^{\circ}$ ($90^{\circ}-b$
is the angle from the $\hat{z}$ axis). To obtain the total number of events that are expected for an exposure of 100 tonne year, one has to integrate over the solid angle and multiply by 365 days. This results in a number of WIMPs above 50 keV of about 74. With the same assumption on the WIMP mass and the WIMP-nucleon cross section, the total number of WIMP events expected from DarkSide-20k for 5 years of data taking including the proper nuclear recoil acceptance as in Ref.~\cite{Aalseth:2017fik}, is about 100 since it includes events below 50 keV, which we conservatively do not include when considering a directional detector. 

The recoil rate is clearly anisotropic \cite{Bozorgnia:2012kq}
and points at coordinates 
$(l=270^{\circ},b=0^{\circ})$
opposite to the direction of the Sun motion throughout the Galaxy.
Since the expected signal in the SHM is rotationally 
symmetric around
the Sun direction, the width of the forward peak is better shown on one dimensional
plot as a
function  of the galactic longitude $l$, obtained integrating over the galactic latitude $b$ (Fig.~\ref{fig:resolutionEth}). The units on the $y$-axis are events/(100 tonne day 180/$\pi$), such that after integration one obtain about $0.2$ events/(100 tonne day). In Fig. \ref{fig:resolutionEth} the effect of different energy thresholds is considered. Indeed, it shows that the width of the peak is slightly reduced for higher recoil energies, thus increasing the correlation between the recoil direction and the apparent WIMP arrival direction, even if the width of the peak is dominated by the WIMP transverse velocity distribution. Indeed, in a liquid the straggling of the recoiling nucleus will broaden further the peak. A higher threshold, in addition, lowers the total rate as it can be quantitatively seen from the normalization factors.

\section{Recoil directional signals at LNGS}
\label{sec:DirectionalRecoilRateLNGS}
In this section  we consider WIMP scattering in a reference frame at rest relative to
a detector situated at LNGS with the $\hat{z}$ axis (anti-parallel to the drift electric field direction) along the local vertical. In this
frame we call the angle between the recoiling nucleus and the vertical axis 
$\theta_r$. In particular we study the expected rates, Eqs. (\ref{eq:RecoilEnergyIntegrated})
and (\ref{eq:FoldedRate}), as function of $\cos\theta_r$ and of the time of the day. The effect
of finite angular resolution is also considered.

\begin{figure}[htbp]
\centering{}\includegraphics[angle=0,width=8cm]{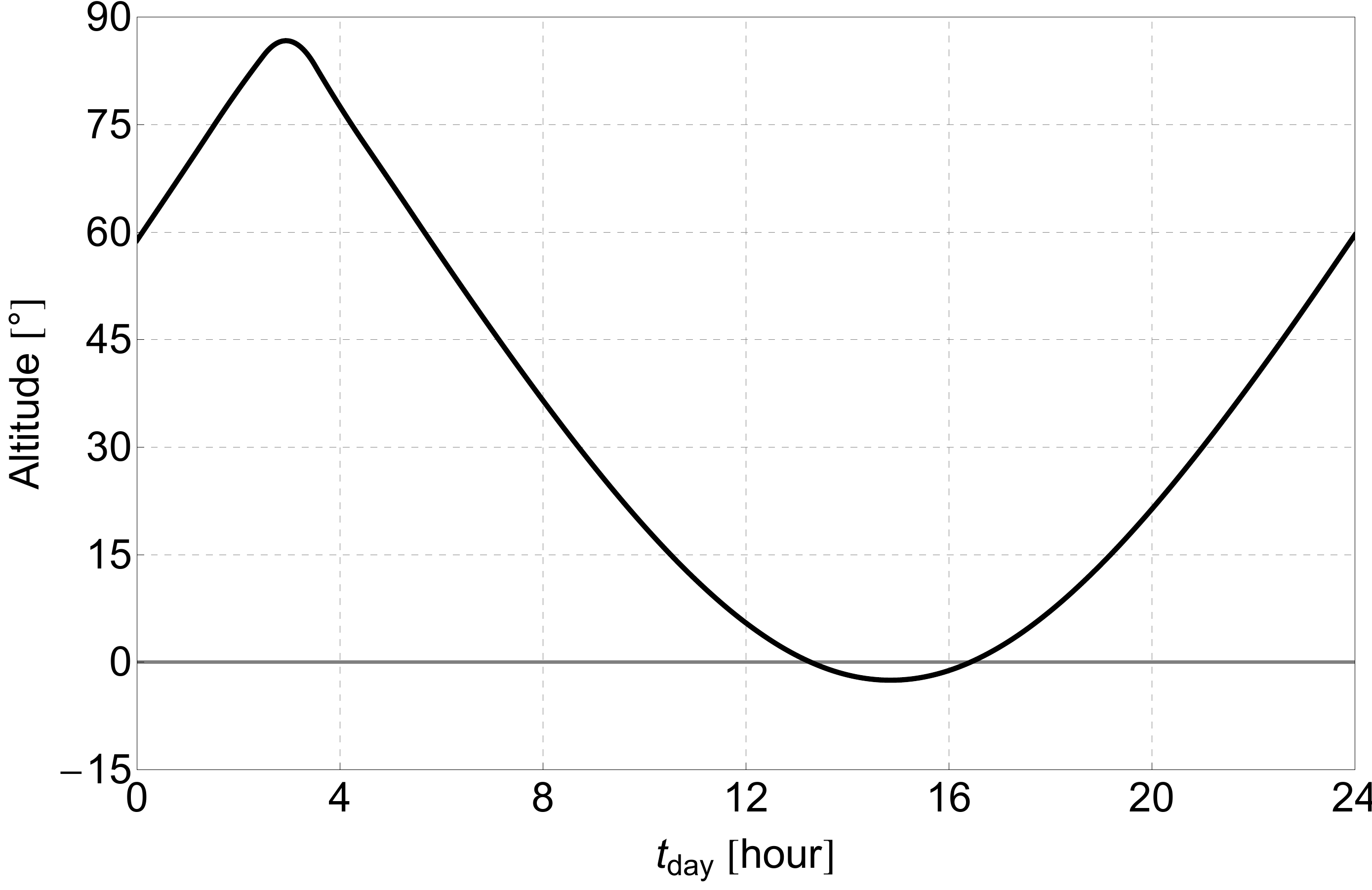} 
\protect\protect\caption[dd]{\label{fig:cygnusPosition} 
Altitude (angle above the horizon)  of
the Cygnus constellation as seen at LNGS as a function
of the local clock time. The horizontal
line at 0\textdegree corresponds to the horizon. }
\end{figure}

Figure \ref{fig:cygnusPosition} shows the Cygnus constellation altitude  
at the LNGS location as a function of the local clock time from the midnight of the Summer Solstice (SS), providing a clear picture of the daily dependence of the expected recoil {\it average} 
direction. As already discussed the correlation between the Cygnus direction
and the WIMP wind changes by at most a tenth of radian during the
year because of the Earth revolution around the Sun.
As the cross section in Eq.~(\ref{eq:DoubleDiffCrossSection})
peaks in the forward direction, when Cygnus is close to the zenith on average 
nuclei recoil mainly towards the nadir, and when Cygnus is close to
horizon on average nuclei recoil mainly in the horizontal plane.
The most important qualitative feature in Fig.~\ref{fig:cygnusPosition}
is that Cygnus spans the whole range of polar directions from zenith to horizon
during the day at the LNGS latitude, thus allowing a strong correlation
between time and polar angle of the recoils. Since
the Cygnus polar angle period is the sidereal day, this correlation is lost
during the year if local solar time is used.

\subsection{Differential rates as functions of the polar angle}

\begin{figure}[htbp]

\centering{}\includegraphics[angle=0,width=8cm]{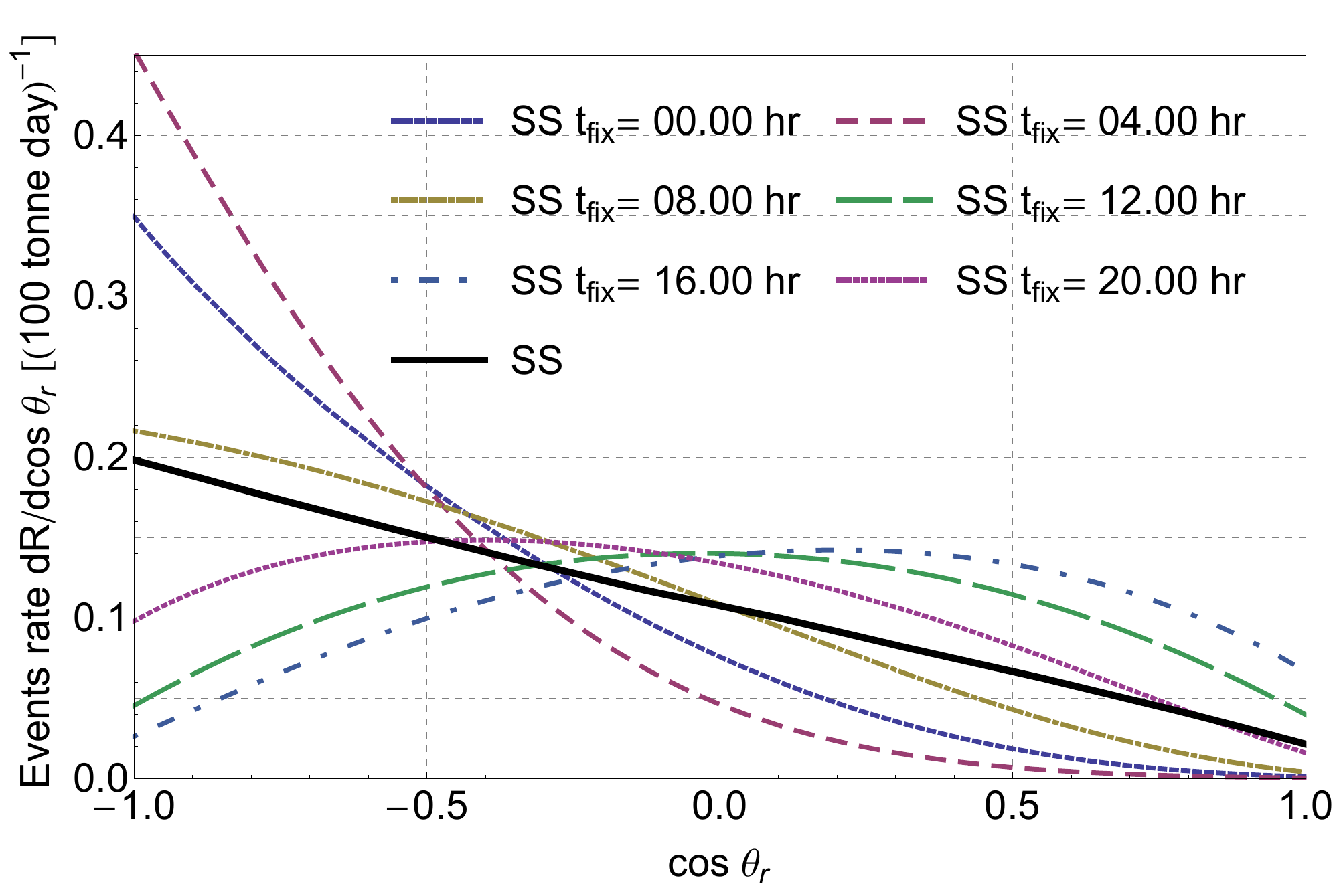}
\protect\protect\caption[ee]{\label{fig:cosTheta}Differential recoil rate as a function of the
cosine of the
polar angle $\theta_{r}$ (the angle between the recoil direction and the z axis) at the latitude of LNGS for the SS day (solid black line). The six dashed curves show the differential recoil rate obtained freezing the position of the Cygnus in the sky at a given time (four-hour apart from each other) of the SS day for the whole day.}
\end{figure}

Figure \ref{fig:cosTheta} shows the differential recoil rate, Eq.
(\ref{eq:polarRate}), as a function of  $\cos\theta_{r}$
for the SS day (solid black line).  This rate is more than twice as high for negative values of $\cos\theta_{r}$ than for positive values, since  Cygnus is most of the time above the horizon. Freezing the position of the Cygnus in the sky at a given time of the SS day for the whole day, one obtains the different dashed lines in Fig. \ref{fig:cosTheta}. One clearly sees that there is a strong dependence on the time of the day. Indeed, the asymmetry in $\cos\theta_{r}$ (the angle between the recoil direction and the z axis) is larger when Cygnus is
high in the sky, \textit{e.g.}, at hour 4, while it is smaller when it is close to the 
horizon, \textit{e.g.}, at hour 16.

\begin{figure}[htbp]

\centering{}\includegraphics[angle=0,width=8cm]{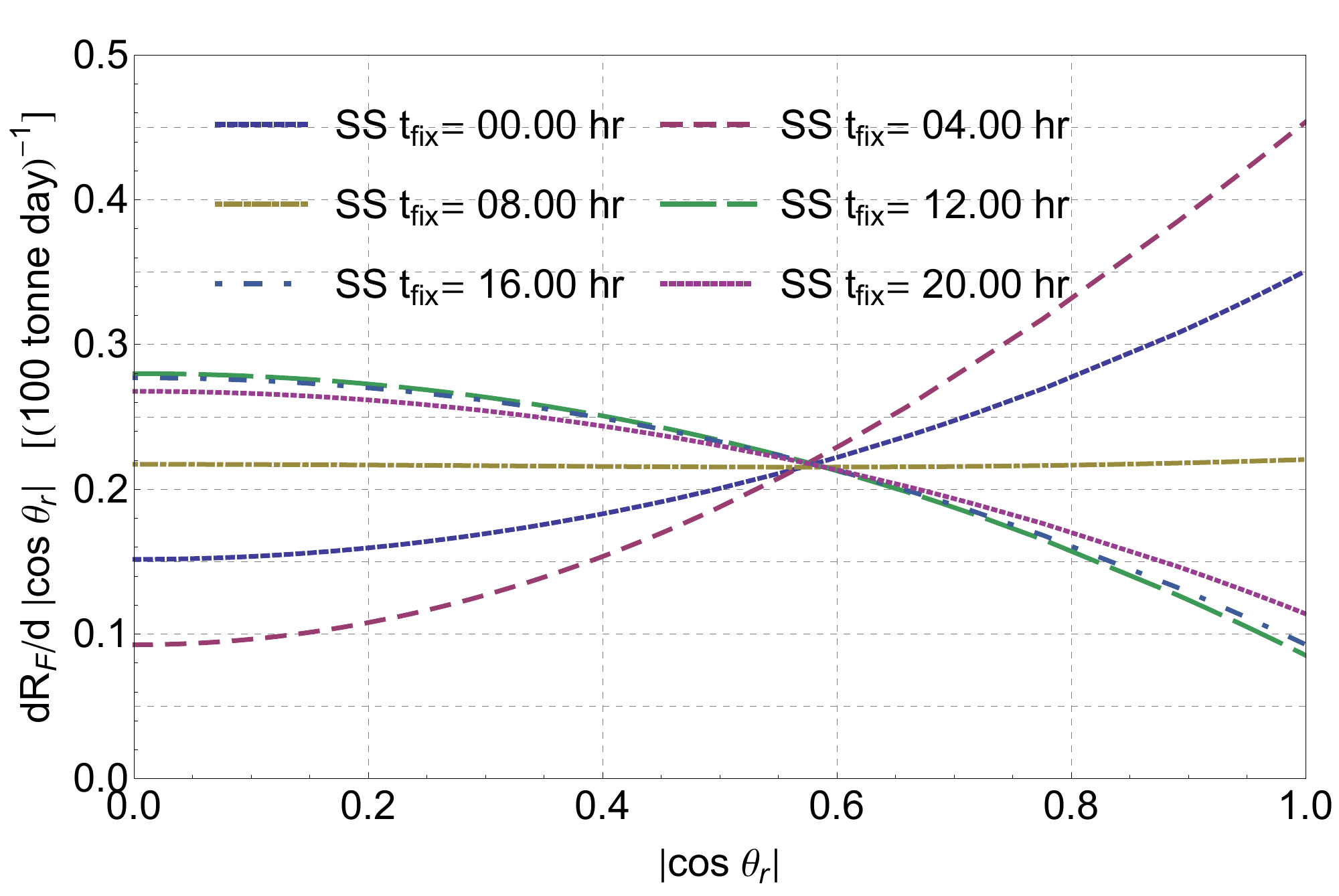}
\protect\protect\caption[ff]{\label{fig:foldedCosTheta} Same as figure~\ref{fig:cosTheta} for
the ``folded'' rate in Eq. (\ref{eq:FoldedRate}).}
\end{figure}

Figure \ref{fig:foldedCosTheta} shows the ``folded'' differential
recoil rate introduced in Eq. (\ref{eq:FoldedRate}), 
the relevant rate for a polar detector. The angular and time dependences
of the rate remain quite strong even without the information on which
side of the track the head is. When Cygnus is close to
the zenith (horizon) the rate is peaked at $|\cos\theta_r|\sim1$
($|\cos\theta_r|\sim0$).

\subsection{Vertical and horizontal event categories}
\label{subsec:VerHor} 

\begin{figure}[htbp]
\centering{}
\includegraphics[angle=0,width=8cm]{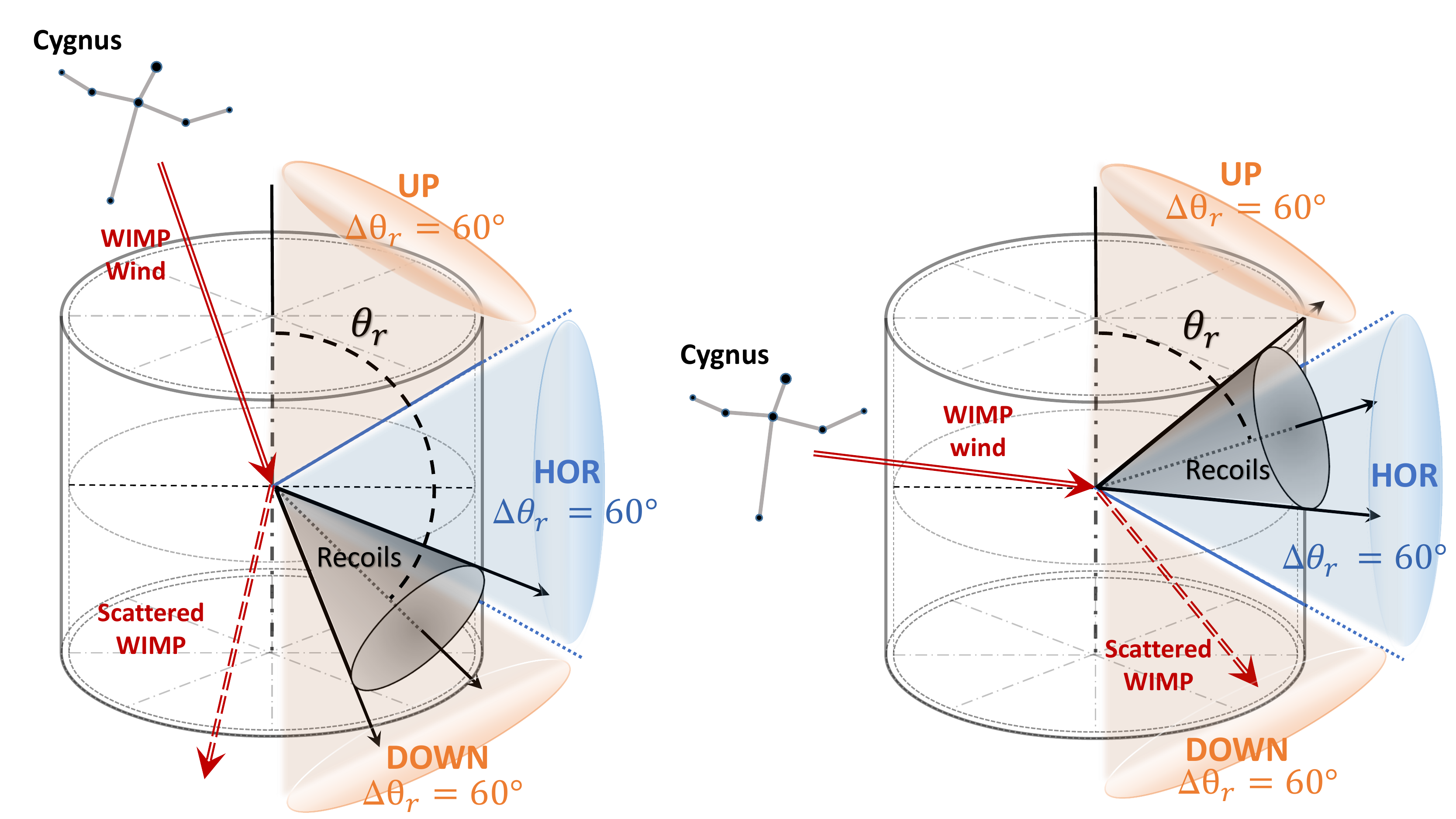}
\protect\protect\caption[ff]{\label{fig:HORvsVERTdef}
Schematic view of the two categories in which events are divided, namely
horizontal (HOR) and vertical (VER=UP+DOWN) events for two different Cygnus altitudes in the sky.
}
\end{figure}

A simple and robust analysis of the time and angular dependency
of the event rate of WIMP collision is achieved by separating
the candidate event sample into two categories that require
only a minimal amount of angular information.
Events can be categorized as horizontal events (HOR), defined by $|\cos\theta_{r}|<0.5$
or $60^{\circ}<\theta_{r}<120^{\circ}$,
and vertical events (VER), defined by $|\cos\theta_{r}|>0.5$
(see Fig.~\ref{fig:HORvsVERTdef}). 

\begin{figure}[htbp]
\centering{}\includegraphics[angle=0,width=8cm]{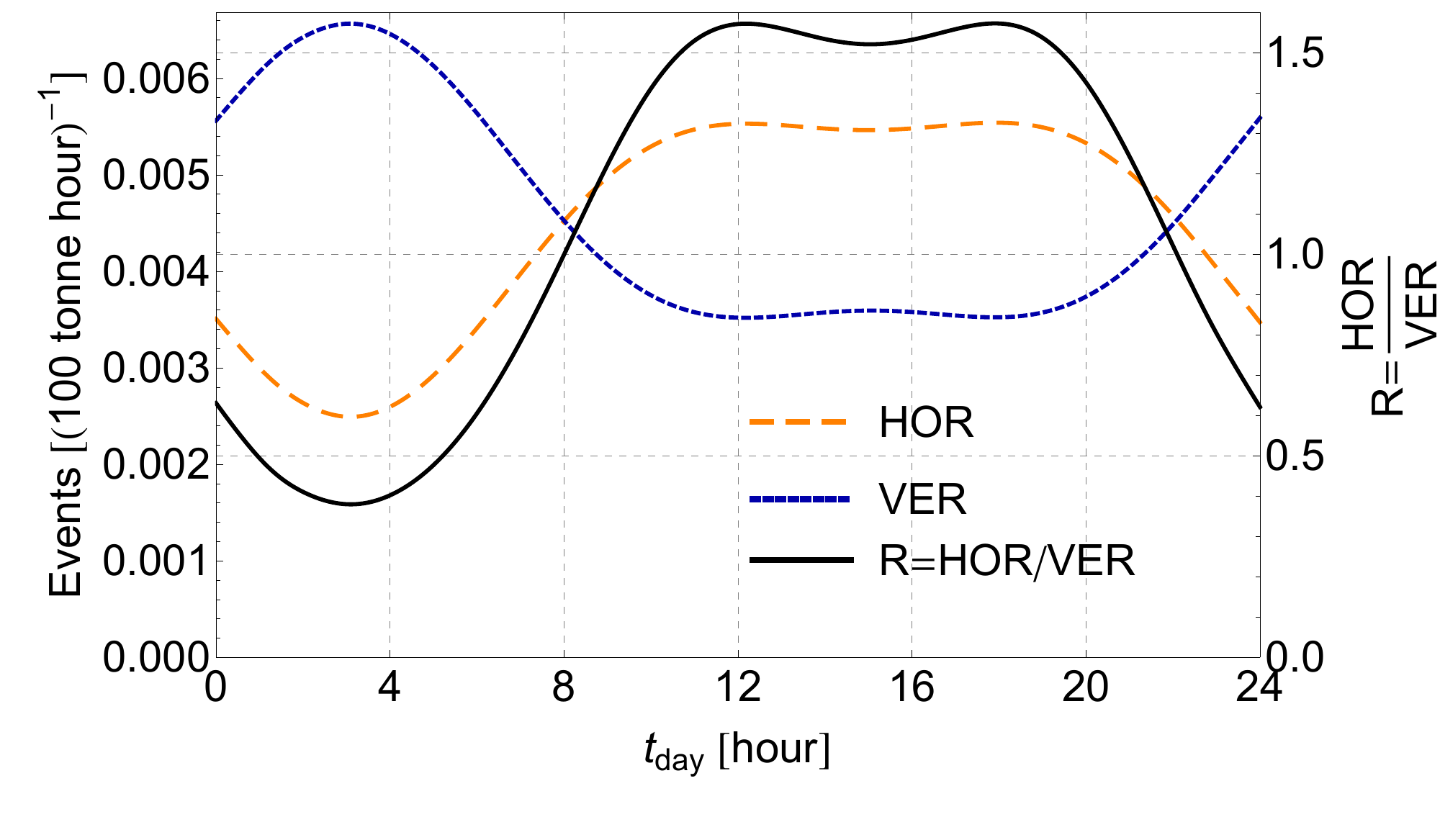}
\protect\protect\caption[gg]{\label{fig:RatioHorVerSS}Horizontal (HOR), 
corresponding to $|\cos\theta_{r}|<0.5$
or $60^{\circ}<\theta_{r}<120^{\circ}$ (long dashes) and vertical
(VER), corresponding to $|\cos\theta_{r}|>0.5$ (short dashes) event
rates as a function of time in event per 100 tonne per hour (left scale).
The solid line shows the ratio $R=\mathrm{HOR}/\mathrm{VER}$ (right
scale). Curves are drawn for the summer solstice day.} 
\end{figure}

Figure \ref{fig:RatioHorVerSS} shows horizontal and vertical WIMP
event rates as function of the time of the day. At the latitude of
LNGS, the time signature of an anisotropic WIMP wind is evident
in spite of the very crude angular classification. In the same figure
we also show the ratio $R=\mathrm{HOR}/\mathrm{VER}$ of horizontal
to vertical events. 
For the given choice of parameters, $R$ changes during the day by a factor of about 4.

\section{Seasonal effects}
\label{sec:SeasonalEffects}

\begin{figure}[htbp]
\begin{centering}
\includegraphics[angle=0,width=8cm]{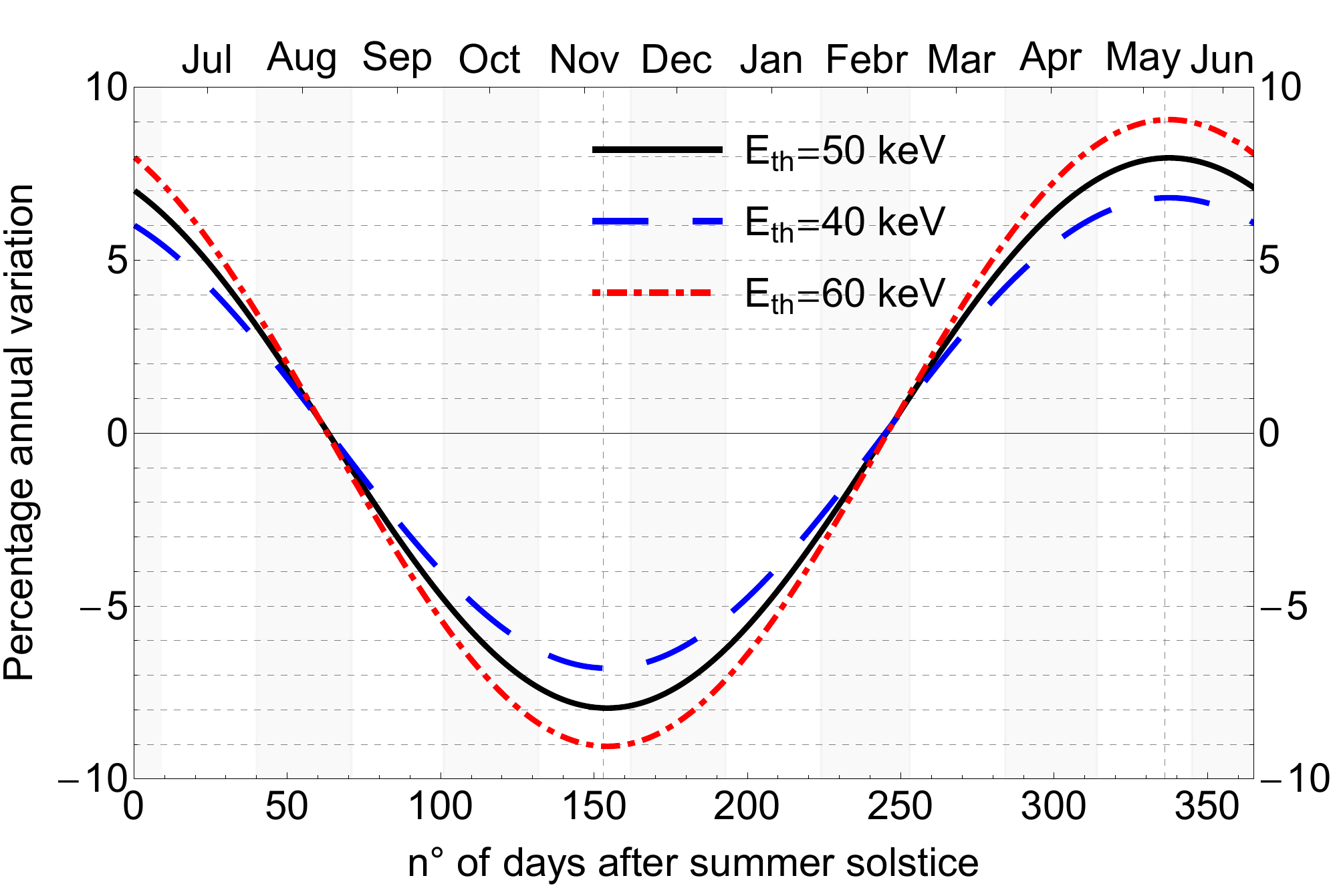}
\par\end{centering}
\protect\caption[jj]{\label{fig:annualPercentualVariationTotalRate}
  Percentage annual variation
of the  argon detector event rate as a function of the number of days
from the summer solstice for three different recoil energy thresholds. The
solid line corresponds to $E_{th}=50$ keV, the dotted line corresponds to $E_{th}=40$ keV, and the dashed
line corresponds to $E_{th}=60$~keV. The corresponding average daily rates are 0.21,
0.27, and 0.15 events per day for a 100 tonne active mass.}
\end{figure}

As already discussed in Section \ref{subsec:Kinematic},
the Earth velocity  within the Galaxy and, therefore, 
the velocity relative to the  
average WIMP velocity  $\mathbf{V}=\mathbf{V}_{SG}+\mathbf{V}_{ES}$
changes during the year due to the annual rotation of orbital velocity
$\mathbf{V}_{ES}$. Since $|\mathbf{V}_{SG}|\approx220$ km/s and
$|\mathbf{V}_{ES}|\approx30$ km/s with an angle of about $60^{\circ}$
between $\mathbf{V}_{SG}$ and the ecliptic, the module $|\mathbf{V}|$ changes
by about $\pm15/220\approx\pm7\%$ during the year causing a similar change of the WIMP flux, while
the annual change of direction is about a tenth of radian. This annual
change of the average WIMP speed produces a corresponding change of
the total daily rate, which reaches its maximum around the end of
May and its minimum around the end of November, as clearly visible
in Fig.~\ref{fig:annualPercentualVariationTotalRate} for three
threshold energies: $E_{th}=40$, 50, and 60~keV. As 
expected~\cite{Lewis:2003bv,Freese:2013gx}
the larger the energy threshold the larger the percentage annual modulation.

\begin{figure}[htbp]
\centering{}\includegraphics[angle=0,width=8cm]{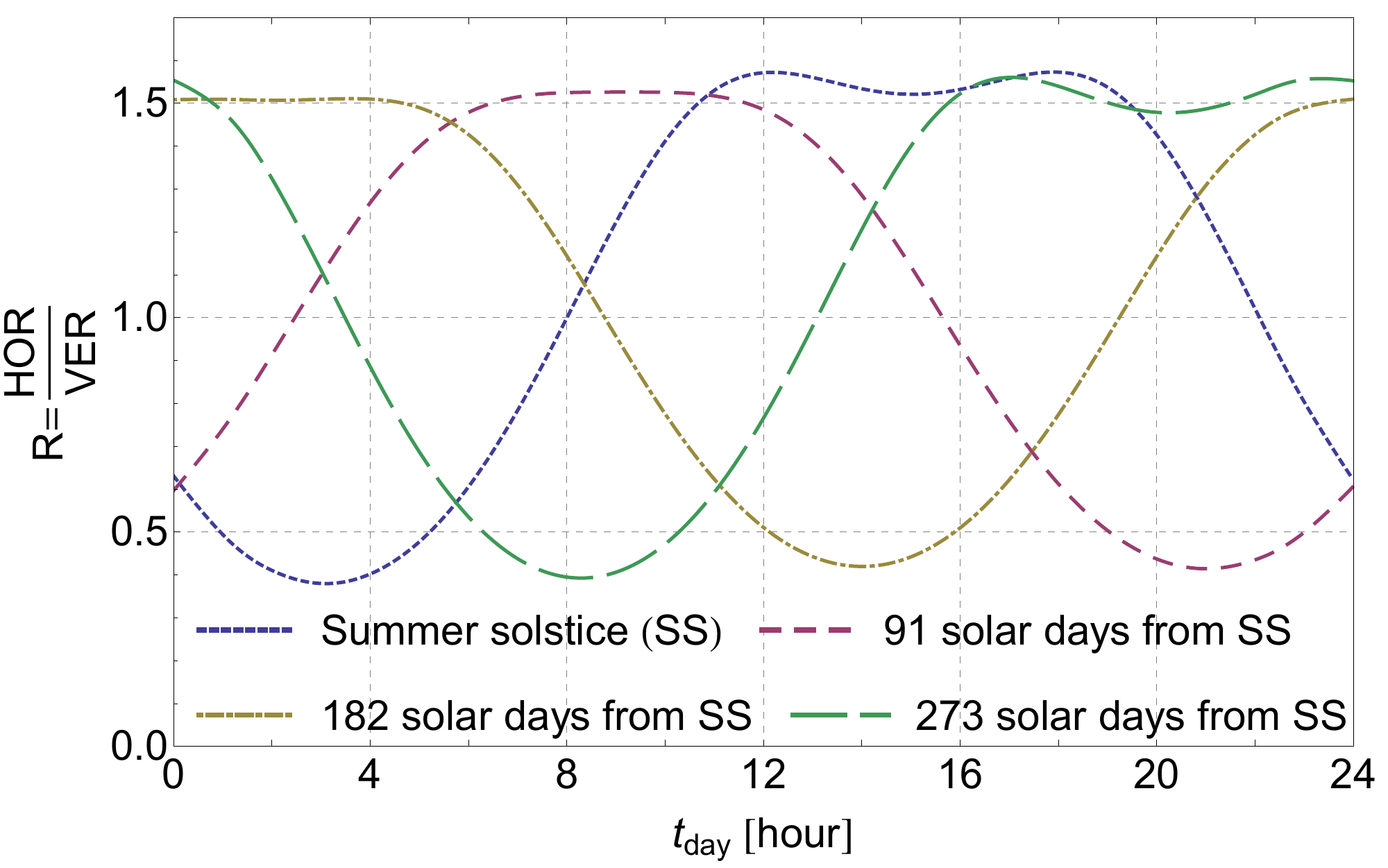} 
\protect\protect\caption[kk]{\label{fig:RatioAnnualDriftSolarDay} Ratio $R$ of expected number
of events along the horizontal and the vertical direction in an argon
detector located at LNGS as function of the time of the day at four
different days of the year. $R$ is defined in Fig. \ref{fig:RatioHorVerSS}
and in the text referring to it.}
\end{figure}

Figure \ref{fig:RatioAnnualDriftSolarDay}
shows the daily
variation of the ratio $R$ at four times of the year;
the signal time
structure changes during the year as function of the local time.
Cygnus, \textit{i.e.} the WIMP direction,
returns exactly in the same position in the sky after a sidereal day,
which is about four minutes shorter than the solar day. This
annual drift of the angular signal as a function of the solar time can
be used to characterize the WIMP signal with respect to other effects that
also produce daily variations but with solar-day periodicity~\cite{Tatarowicz:2011bn}.

\begin{figure}[htbp]
\centering{} \includegraphics[angle=0,width=8cm]{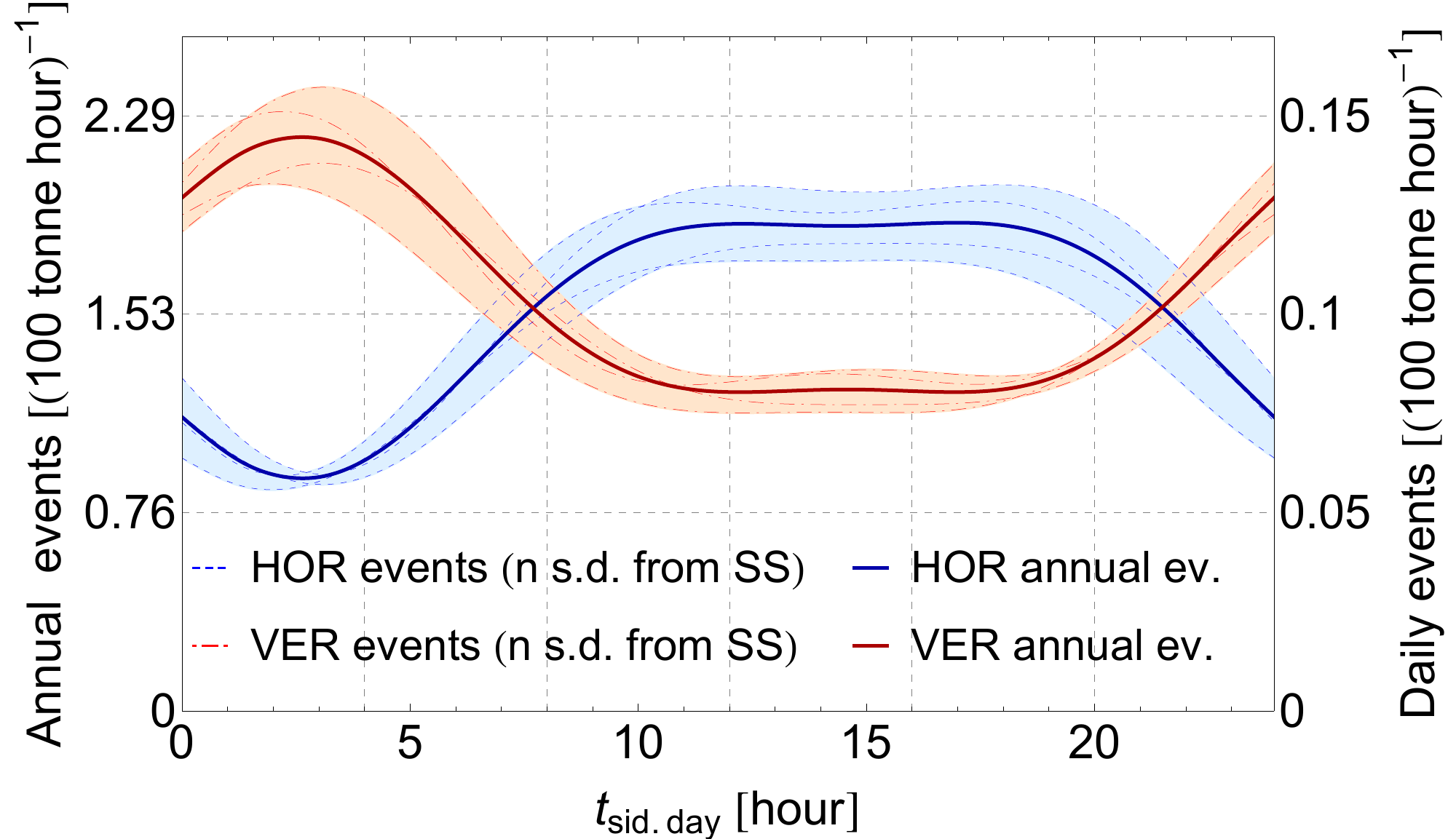}
\protect\protect\caption[ll]{\label{fig:averageHorVerAnnaul} Solid red and blue lines (left y-scale): annual sum of HOR $\equiv|\cos\theta_{r}|<0.5$) and vertical (VER $\equiv|\cos\theta_{r}|>0.5$) events expected at each hour of a sidereal day, respectively. Dashed red and blue lines(right y-scale): events at each hour of a given sidereal day, namely the day of the SS, and 91, 182, 273 days after it.
}
\end{figure}

\begin{figure}[htbp]
\centering{} \includegraphics[angle=0,width=8cm]{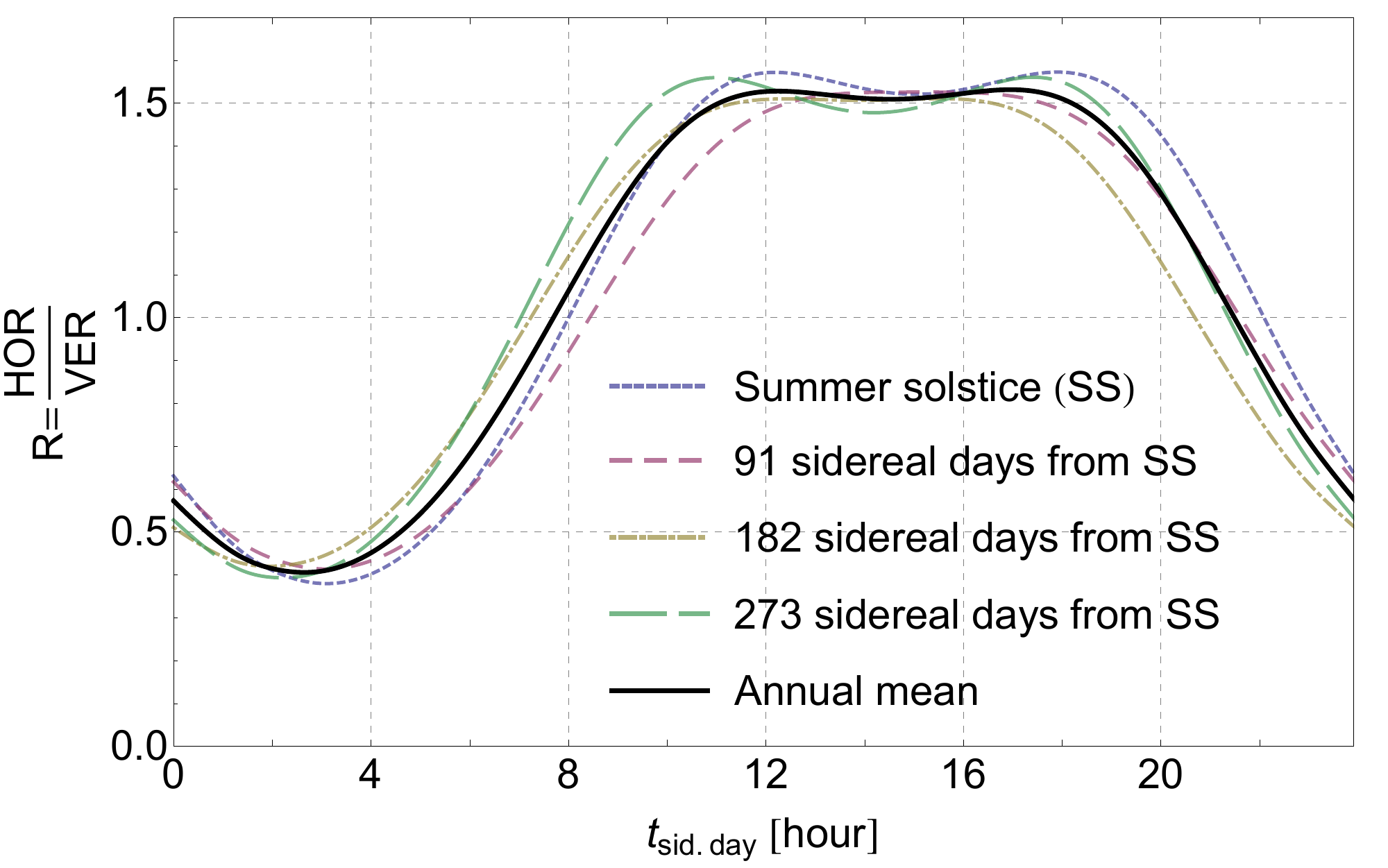}
\protect\protect\caption[mm]{\label{fig:averageRatioAnnual} 
Same as Figure Fig.~\ref{fig:averageHorVerAnnaul}
for the ratio $R=\mathrm{HOR}/\mathrm{VER}$.}
\end{figure}

If the sidereal time is used to time-stamp events, the annual drift is
eliminated and it is possible to compare and average events rates
taken at different days. Figure {\ref{fig:averageHorVerAnnaul} shows
the HOR and VER daily event rates taken at four
times of the year and the annual rates of HOR and VER events computed by summing the contributions at each sidereal day. Figure \ref{fig:averageRatioAnnual} shows for $R$ the same information
as Fig. \ref{fig:averageHorVerAnnaul} for HOR and VER. Note that
part of the seasonal variation of HOR and VER cancels out in their
ratio $R$.

The much larger time variation of the directional signal relative to the seasonal variation of the 
non-directional signal is evident by
comparing Fig.~\ref{fig:annualPercentualVariationTotalRate} 
to Fig. \ref{fig:averageRatioAnnual}, or also in
Fig. \ref{fig:RatioAnnualDriftSolarDay} or Fig. \ref{fig:averageHorVerAnnaul}. However one should use the combination of seasonal and directional modulations to better characterize the nature of the signal.

\section{Statistical analysis for a directional signal}
\label{sec:statanal}

A path through the discovery process of DM searches may proceed initially through
the observation of a number of candidates that significantly
exceed the small expected background level. However, after rejecting the background-only
hypothesis, the study of angular properties of the observed nuclear recoils can corroborate
the belief that the observed signal can be attributed to DM interactions. 
In Sec. \ref{sec:DirectionalRecoilRateLNGS} we have discussed 
semi-quantitatively the power of angular discrimination by using a crude classification 
in horizontal and vertical events. Here  we want to quantitatively discuss the number of events 
necessary to discriminate the 
hypothesis of a DM signal with preferential incoming direction from the Cygnus constellation
against the alternative hypothesis of an isotropic signal.


The negative logarithm of the likelihood ratio is taken as test statistic $t$
to discriminate between the hypotheses of a directional signal from the Cygnus constellation (Cyg)
against an isotropic signal (iso) ($\mathbf{V}$ as in Section~\ref{subsec:Kinematic} or $\mathbf{V}=0$, respectively).
Such test statistic can also be extended in order to take into account the 
effect of systematic uncertainties for realistic applications \cite{Cowan:2011cx}. The test statistic is defined as 
\begin{equation}
\label{eq:testStat}
  t(\vec{x}^{(1)},\cdots,\vec{x}^{(N)}) =
  -\ln\frac{{\cal L}_{\mathrm{Cyg}}(\vec{x}^{(i)})}{{\cal L}_{\mathrm{iso}}(\vec{x}^{(i)})}\,,
\end{equation}
where $ {\cal L}_{\mathrm{Cyg},\mathrm{iso}}(\vec{x}^{(i)}) $ are the likelihood functions corresponding to the 
two hypotheses. 
Given a sample of $N$ independent WIMP events, the two likelihood functions are given by the products 
of the probability density function (PDF)  values $f_{\mathrm{Cyg},\mathrm{iso}}(\vec{x}^{(i)})$ 
corresponding to each WIMP interaction candidate:
\begin{equation}
  {\cal L}_{\mathrm{Cyg},\mathrm{iso}}(\vec{x}^{(1)},\cdots,\vec{x}^{(N)}) =
  \prod_{i=1}^{N}f_{\mathrm{Cyg},\mathrm{iso}}(\vec{x}^{(i)})\,,
\end{equation}
where 
 the vector $\vec{x}^{(i)}$ contains the variables used to characterize the event  $(i)$. 
In the present work we use the two variables 
$\theta_\mathrm{rec}$, the recoil polar angle in the laboratory, and 
$\theta_{\mathrm{Cyg}}$, the polar angle 
of the Cygnus constellation at time of the event in the laboratory:
 $\vec{x}^{(i)} \equiv (\theta_\mathrm{rec}^{(i)}, \theta_{\mathrm{Cyg}}^{(i)})$.
Additional variables such as the recoil energy or
the time of the year  could provide additional information and, in principle, better discrimination
between the two hypotheses, our conclusions are conservative in this
respect. The same method can be used  to study alternative models for WIMP distribution or 
backgrounds.

The PDFs $f_{\mathrm{Cyg},\mathrm{iso}}(\vec{x}^{(i)})$ have been sampled generating
$10^{10}$ simulated interaction recoils for each hypothesis and 
binning the allowed kinematic range of each variable $\theta_\mathrm{rec}^{(i)}$ and $ \theta_{\mathrm{Cyg}}^{(i)}$
with 100 bins. In the simulation the
energy has been smeared by 10~keV in order to account for the energy resolution and
an energy threshold of 50~keV has been used. In addition we compared the case of perfect
resolution of the recoil angle in the laboratory frame to a resolution 
smeared by a Gaussian distribution with a
400~mrad width.

Given an assumed number $N$ of WIMP interaction candidate events, $10^{7}$ 
pseudo-samples of $N$ events
each were generated for each of the four cases, namely events from Cygnus direction or isotropic and with the two angular resolutions.
The test statistic, $t$, of Eq.~(\ref{eq:testStat}) has been evaluated for each pseudo-sample and stored into 
histograms with a fine binning.
Figure~\ref{fig:likRatio} shows the distribution of the test statistic $t$ defined in 
Eq.~(\ref{eq:testStat}) for the case of ideal  (left panel) and 400~mrad resolution (right panel) with $N=50$. The directional
(isotropic) distribution is peaked at negative (positive) values.
\begin{figure}[htbp]
\centering{}
\includegraphics[angle=0,width=6cm]{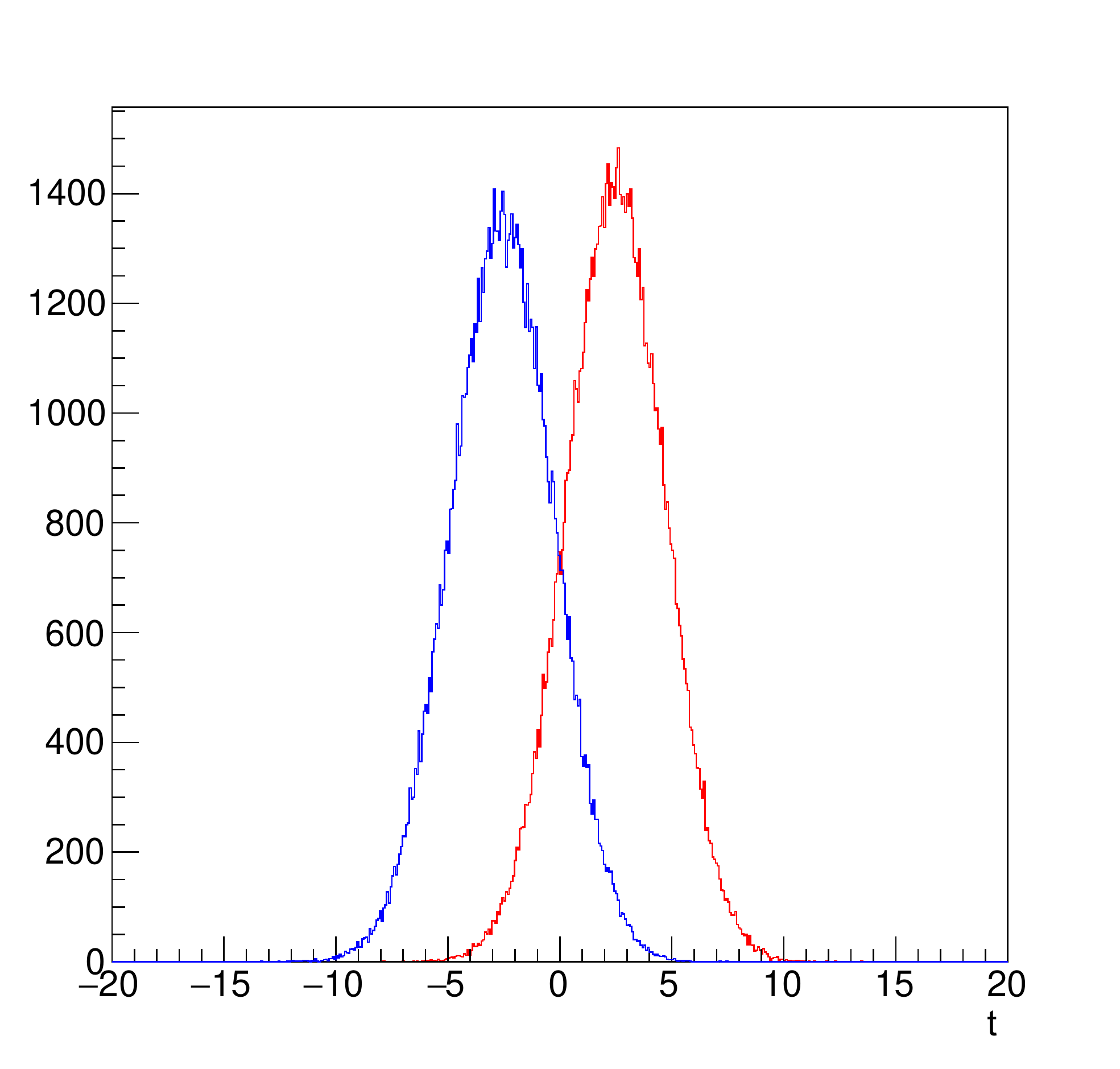}
\includegraphics[angle=0,width=6cm]{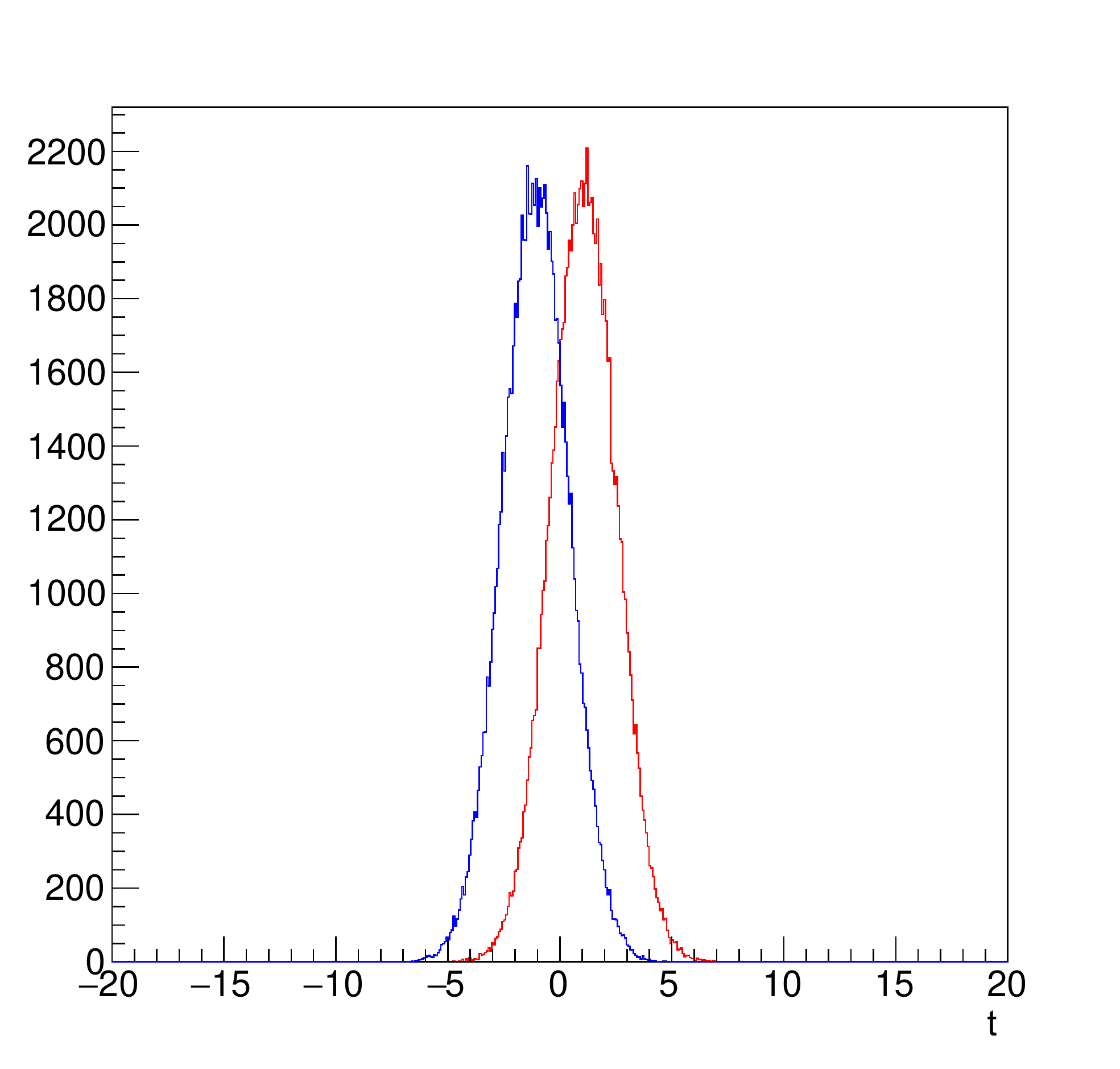}
\protect\protect\caption[ff]{\label{fig:likRatio}
  Distribution of the likelihood-ratio test statistic $t$ for ideal resolution
  (left) and for a resolution of 400~mrad (right) for $N=50$ observed DM
  candidates. The blue (red) curves that peak at negative (positive) values of $t$ correspond to
  the hypothesis of incoming particles from the Cygnus direction (of isotropic signal).
}
\end{figure}
The expected $p$-value is computed from the distribution of the test statistic $t_\mathrm{iso}$
corresponding to the null (isotropic) hypothesis by  considering the percentage of pseudo-sample with $t$ below
$t_{\mathrm{Cyg},0}$, where $t_{\mathrm{Cyg},0}$ is the median of the distribution of the  test statistic $t_{\mathrm{Cyg}}$
corresponding to Cygnus direction hypothesis. The corresponding  one- or two-standard-deviation excursions
are calculated by considering instead of $t_{\mathrm{Cyg},0}$ the boundaries of the one- or two-standard-deviation
interval for the  test statistic $t_{\mathrm{Cyg}}$. 

The expected $p$-values as a function
of the observed number of DM interaction candidate events
are shown in Fig.~\ref{fig:pvalPlots} for ideal angular resolution (left) and for a
400~mrad resolution (right).
\begin{figure}[h]
\centering{}
\includegraphics[angle=0,width=6cm]{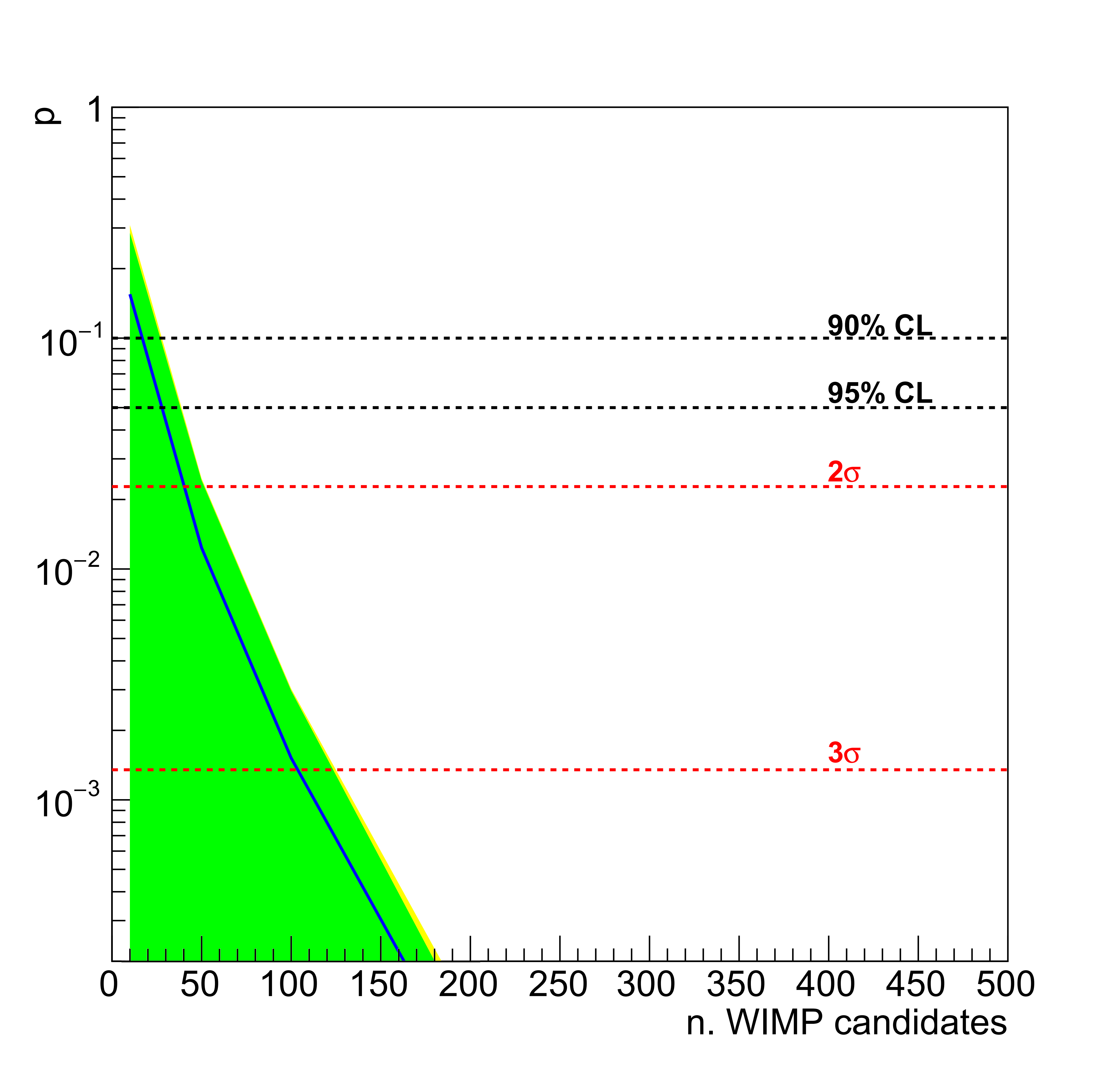}
\includegraphics[angle=0,width=6cm]{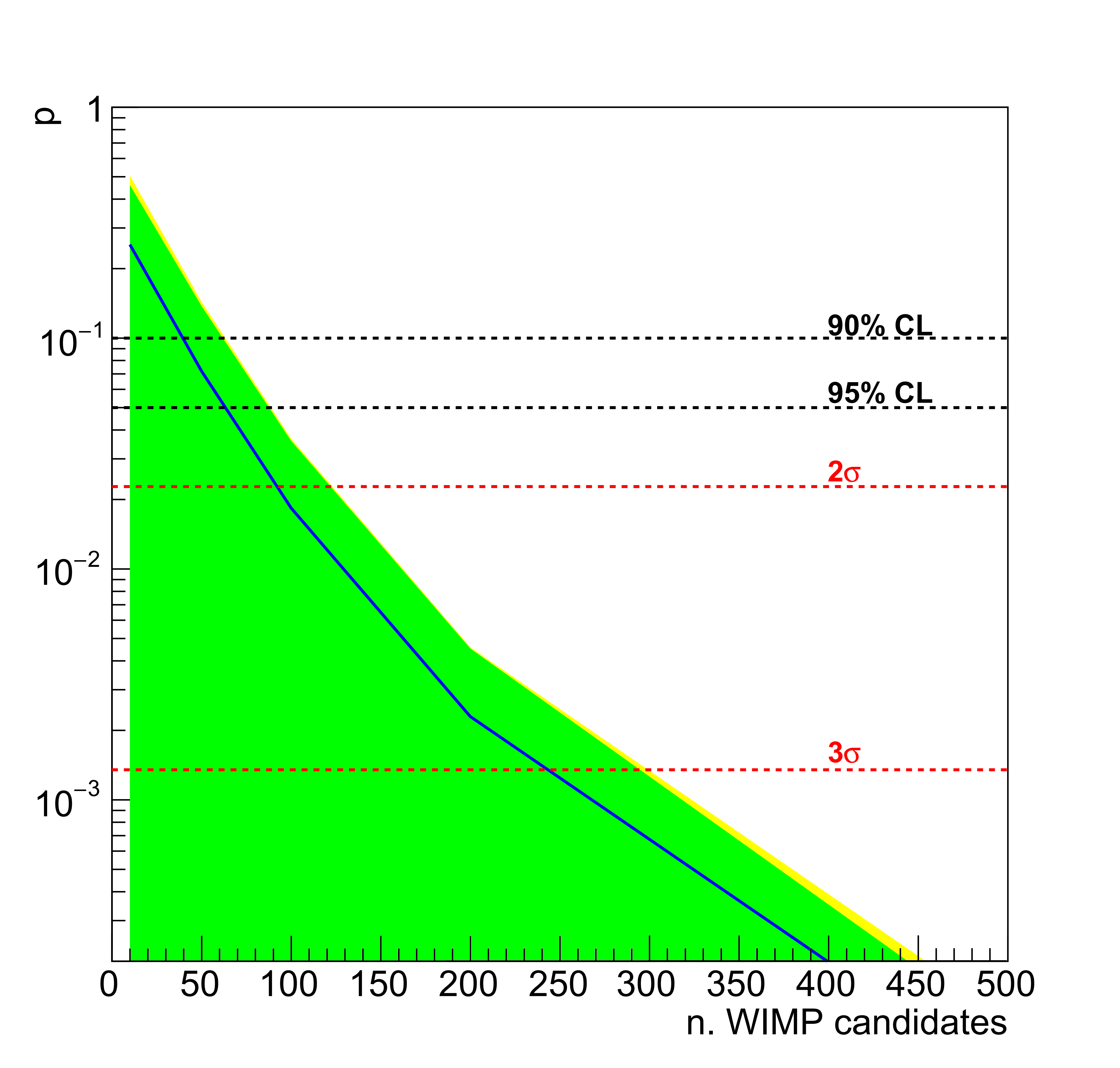}
\protect\protect\caption[ff]{\label{fig:pvalPlots}
  Expected $p$-value (blue line) of the null hypothesis (isotropic signal)
  as a function of
  the observed number of DM interaction candidate events
  for ideal resolution (left) and for a resolution of 400~mrad (right).
  The green and yellow bands show the excursion range at one and two
  standard deviations. The horizontal lines show from top to bottom the 90\% and 95\%
  CL exclusion and the $2\sigma$ and $3\sigma$ significance levels.
}
\end{figure}
In the case of an ideal resolution, a $3\sigma$ evidence of a directional
signal is expected to be achieved with about 100 candidate events.  
For an angular resolution of 400~mrad, a $3\sigma$ evidence can be achieved
with about 250 events.

\section{Conclusions}
\label{sec:conclusions} 

We examined the sensitivity of a large scale dark matter liquid argon experiment to 
the directionality of the dark matter signal, under  the assumption that it is possible, 
above a certain threshold, to measure the direction of the recoiling nucleus. 
This may indeed be possible with two-phase liquid argon detectors, if the suggested
dependence of columnar recombination on the alignment of the recoil momentum with the electric field
can be  experimentally demonstrated. \\
In this paper we study differential rates as a function of the nuclear recoil direction angle 
with respect to the vertical with no head-tail discrimination for a detector located at the 
Laboratori Nazionali del Gran Sasso and diurnal and seasonal modulations of such a signal. 

With a likelihood-ratio based statistical approach we show that, using the angular information alone, 
100 (250) events are sufficient to reject the isotropic hypothesis at 3 sigma level for a perfect (400~mrad) angular resolution. 
For an exposure of 100 tonne years, such as the detector described in~\cite{Aalseth:2017fik}, this 
number of events
corresponds to a WIMP-nucleon cross section of 
about $\textrm{10}^{-46}\textrm{\ cm}^{2}$ at 200 GeV WIMP mass.
Larger exposures  would  probe directionality at smaller cross sections.\\ In view of the evidence presented in this paper, and in consideration of the 
strong exclusion bounds already achieved by null observations performed by non-directional 
dark matter detectors, it is of utmost importance the development of experimental 
technologies able to couple directional sensitivity with large fiducial masses (many tonnes) 
and the ability to collect large exposures free of background from $\beta/\gamma$ events and 
neutron-induced nuclear recoils.  One possible avenue would be offered by the presence of 
the signature of columnar recombination in nuclear recoils in a liquid argon time projection
chamber, where this effect has already been observed for $\alpha$ particles and protons. 
Dedicated experiments performed on monochromatic, pulsed neutron beams will allow to explore the 
possible presence of this signature.

\bibliographystyle{JHEP}
\bibliography{BIBLIO}

\end{document}